\documentclass[%
 reprint,
 amsmath,amssymb,
 aps,
]{revtex4-2}

\usepackage{graphicx}
\usepackage{dcolumn}
\usepackage{bm}
\usepackage{wrapfig} 
\usepackage{xcolor}

\begin{document}

\preprint{APS/123-QED}

\title{Structural States of Filamentary Microgravity Dusty Plasma}

\author{Emerson Gehr\textsuperscript{1}, Abbie Terrell\textsuperscript{1}, Katrina Vermillion\textsuperscript{1}, Alexandria Mendoza\textsuperscript{1}, Bradley Andrew\textsuperscript{2}, Diana Jimenez Marti \textsuperscript{1}, Evdokiya Kostadinova\textsuperscript{2}, Peter Hartmann\textsuperscript{1,3}, Lorin Matthews\textsuperscript{1}, Truell Hyde\textsuperscript{1}}

\affiliation{%
\textsuperscript{1}Center for Astrophysics, Space Physics, and Engineering Research and Department of Physics, Baylor University, TX, USA  
\\ \textsuperscript{2}Department of Physics, Auburn University, AL, USA 
\\ \textsuperscript{3}Institute for Solid State Physics and Optics, Wigner Research Centre for Physics, P.O. Box 49,
H-1525 Budapest, Hungary
}%

\date{\today}

\begin{abstract}
This study investigates the filamentary structural states of microgravity dusty plasma using data from the Plasmakristall-4 (PK-4) facility on board the International Space Station. The dust particles in the PK-4 discharge are observed to form field-aligned filaments and nested (layered) structures in response to changes in the plasma conditions, neutral gas pressure, and externally applied electric field. This work explores the possibility that these filamentary dusty plasmas exhibit properties of liquid crystals. The structural characteristics of the dust clouds are studied for nine sets of pressure-current conditions using pair correlation functions calculated for particles (i) within individual filaments, (ii) within the central plane of the dust cloud, and (iii) within successive planes of the cloud (the 3D cloud). It is observed that, at low pressure ($\approx$30 Pa), the entire cloud is in a weakly crystalline state with similar coupling of particles within filaments and among neighboring filaments. At high pressure ($\approx$70 Pa), the order within filaments improves (enhanced crystalline behavior), while the degree of freedom of filaments to move with respect to each other increases (enhanced liquid behavior). Since neutral gas pressure in dusty plasma acts as inverse temperature, we argue that the structural changes observed with increasing pressure are analogous to a transition to a nematic liquid crystal state. It is further observed that the filaments exhibit alignment in nested surfaces for several pressure-current conditions, suggesting the possibility of a smectic liquid crystal state. These results are confirmed by molecular dynamics simulations of the dust and ions using the DRIAD (Dynamic Response of Ions And Dust) code. 
\end{abstract}

\maketitle


\section{\label{sec:intro}Introduction}

\begin{figure}
    \centering
    \includegraphics[width=0.45\textwidth]{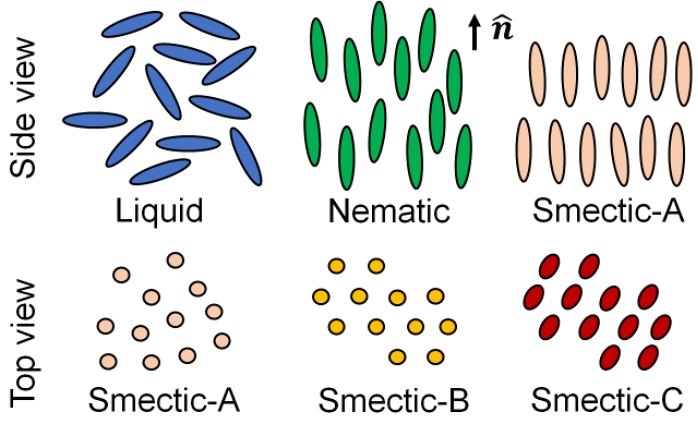}
    \caption{Patterns and order observed in LC phases}
    \label{fig:lcphases}
\end{figure}

Materials in the liquid crystal (LC) state of matter exhibit properties of both conventional liquids and solids. Beyond their well-known use in LC displays \cite{chen_liquid_2018}, \cite{white_new_1974}, liquid crystals can potentially transform technology and medicine through the development of soft robotics \cite{lv_stimulus-driven_2021}, \cite{shen_stimuli-responsive_2020}, artificial muscles \cite{li_artificial_2006}, \cite{buguin_micro-actuators_2006}, and LC polymers \cite{liu_liquid-crystalline_2021}, \cite{collyer_liquid_1993}, \cite{woltman_liquid-crystal_2007}. Liquid crystals are also known to exhibit the electrorheological effect (change of viscosity due to electric-field-induced structural transition) \cite{hao_electrorheological_2001}, \cite{hao_positive_2005}, \cite{tao_electric-field-induced_1993}, which expands their application to the development of fast clutches and hydraulic valves, brakes, and shock absorbers \cite{sadeghi_innovative_2012}, \cite{choi_speed_2007}, \cite{stanway_applications_1996}. The successful utilization of LCs in all these applications relies on controlled self-assembly and transitions into different LC phases, illustrated in Fig.~\ref{fig:lcphases}. 

In the liquid-nematic phase transition, the LC molecules align along a preferred direction (indicated by $\hat{n}$ in Fig.~\ref{fig:lcphases}), while in the nematic-smectic-A transition, the molecules additionally arrange in layers. In other smectic phases, like the smectic-B and smectic-C phases shown in Fig.~\ref{fig:lcphases}, hexagonal patterns form in the direction perpendicular to $\hat{n}$. Understanding and controlling pattern formation in various nematic-smectic transitions will enable promising methods for fast and affordable fabrication of optical microarrays \cite{kim_optically_2010}, \cite{son_optical_2014} and assembly of functional nanoparticles \cite{coursault_tailoring_2015}, \cite{gryn_electric-field-controlled_2016}. One of the outstanding open questions in the study of liquid crystals is the universality of the nematic-smectic-A transition \cite{zappone_analogy_2020}, \cite{yethiraj_recent_2007}. In addition, a clear connection has yet to be established between pattern formation and the critical behavior of LCs at the various nematic-smectic transitions \cite{zappone_analogy_2020}. Microscopic experimental treatment of liquid crystals is challenging due to their high material density, the presence of long-range interactions, and many-body effects. The main goal of the present work is to explore the possibility of studying LC phenomena using microgravity complex (dusty) plasma analogue systems, which consist of charged micron-sized particles suspended in weakly-ionized gas.

Dust grains immersed in plasma typically become negatively charged and are subject to both ion drag forces and collective interactions. As the magnitude of the dust charge depends on both the dust location within the plasma (e.g., plasma sheath vs. bulk plasma) and on the time evolution of the plasma conditions, there can be a wide variation in the coupling strength among dust particles in complex plasma, resulting in the formation of diverse structural states. Dusty plasma have been shown to self-organize into strongly coupled fluids and crystalline structures \cite{chu_direct_1994}, \cite{thomas_plasma_1994}, \cite{nefedov_formation_2001}, \cite{hartmann_crystallization_2010}, \cite{joshi_recrystallization_2023}, which makes them ideal for the study of self-organization and stability, phase transitions, and transport phenomena. The main advantage of using a dusty plasma analogue to study fundamentals of materials science phenomena is the ability to observe the dust particles at the kinetic level with a video camera, thus, allowing for the reconstruction of the entire phase space.

On Earth, the liquid crystal properties of dusty plasma have been demonstrated in experiments with elongated dust particles in capacitively coupled RF discharges \cite{molotkov_liquid_2000}, \cite{banu_precession_2015} and cylindrical dust particles in DC discharges \cite{fortov_dusty_2002}, \cite{khrapak_liquid_1999}. In microgravity, dusty plasma has been shown to exhibit electrorheological (ER) effects, thus demonstrating its potential use as an analogue system for the study of phase transitions in ‘smart’ materials \cite{ivlev_first_2008}, \cite{ivlev_electrorheological_2010}, \cite{dietz_phase_2021}. However, unlike conventional electrorheological fluids, where the transition to a filamentary state is induced by polarization of the grains themselves, effective polarization in electrorheological dusty plasma is caused by the dust interacting with an anisotropic ion flow. Specifically, streaming ions are deflected and focused by a negatively charged dust grain, which results in the formation of ion wakefield downstream from the grain. As a result, the interparticle interaction potential is highly sensitive to plasma conditions, including electric field, pressure, and plasma dynamics. 

The first ER complex plasma was observed in the Plasmakristall-3 (PK-3) Plus laboratory \cite{thomas_complex_2008} on board the International Space Station (ISS). The interaction potential guiding the ER effect in PK-3 Plus was proposed to have a dipole-dipole form, which seemed to yield agreement between molecular dynamics (MD) simulations and experiments \cite{ivlev_electrorheological_2010}. Here we investigate the filamentary structural states of microgravity dusty plasma using data from the PK-3 successor - the Plasmakristall-4 (PK-4) facility, currently on board the ISS \cite{pustylnik_plasmakristall-4_2016}. After the first observation of filamentary structures in the PK-4 DC discharge, it was conjectured that the interparticle interaction has a similar dipole-dipole form as in PK-3 Plus \cite{ivlev_electrorheological_2010}, \cite{kompaneets_design_2009}, \cite{ivlev_complex_2011}. However, recent numerical simulations \cite{vermillion_influence_2022}, \cite{matthews_effect_2021} suggest that the formation of filamentary structures in PK-4 is guided by a more complex interaction potential resulting from the interplay between the polarity switched DC electric field, the neutral gas pressure, and the onset of high-frequency ionization waves present in the PK-4 discharge \cite{hartmann_ionization_2020}. 

In addition to the electrorheological effect, several studies of PK-4 ISS data have shown that the PK-4 field-aligned filamentary structures exhibit simultaneous liquid and crystalline properties in a broad range of experimental conditions \cite{kostadinova_nematic_nodate}, \cite{mitic_long-term_2021}, \cite{pustylnik_three-dimensional_2020}, \cite{yaroshenko_possible_2021}. Parabolic flight experiments and molecular dynamics simulations of dusty plasma in the PK-4 have confirmed that dust filaments can also arrange in hexagonal patterns in the plane perpendicular to the external electric field and form distinct layers \cite{takahashi_study_2014}, \cite{totsuji_study_2014} - characteristic properties of smectic LC phases. Thus, it can be argued that the PK-4 microgravity dusty plasma exhibits unique structural states that combine properties of both electrorheological fluids and liquid crystals. These properties depend on the anisotropic potential arising from the interaction of the charged dust grains with the ion wakefield under different discharge conditions.

\begin{figure}[hbt!]
\centering
\includegraphics[width=0.4\textwidth]{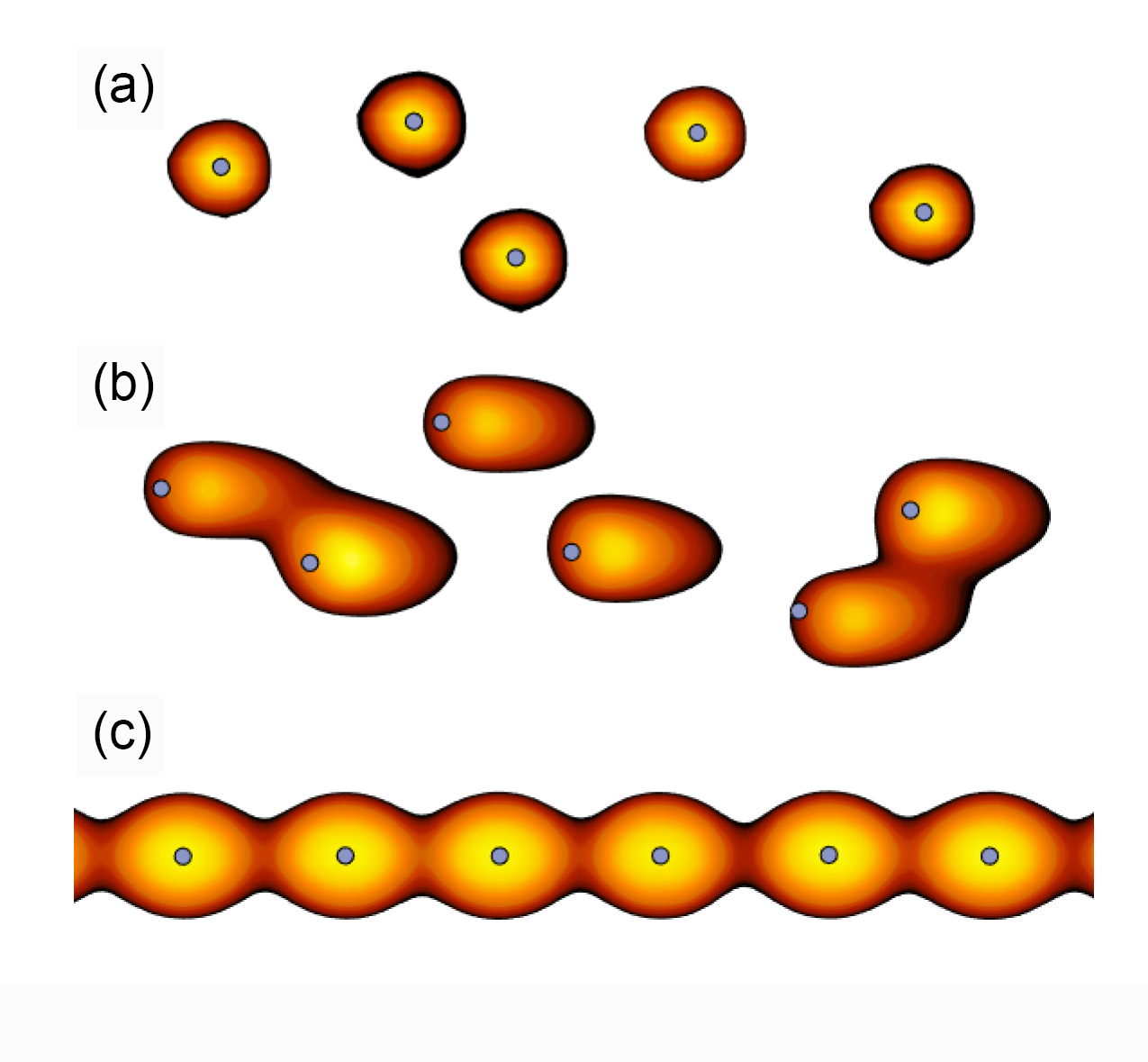}
\caption{Examples of ion accumulation near charged dust grains in the presence of (a) no axial electric field, (b) a DC field (to the right), and (c) a DC field with symmetric polarity switching. The largest ion density is indicated by the lightest shade (yellow) and the lowest ion density is represented by the darkest shade (black). Small gray circles represent the charged dust grains. Ion densities calculated using the DRIAD simulation \cite{vermillion_influence_2022}.}
\label{fig:ionwake}
\end{figure}

As the DC current in PK-4 is increased from zero, the induced electric field causes the dusty plasma cloud to transition from an isotropic (liquid-like) state to a state where the dust particles start aligning along the direction of the electric field due to the formation of ion wakefields downstream from each grain (Fig.~\ref{fig:ionwake}a, b). Filament formation is enhanced by switching the polarity of the electric field with a frequency higher than the dust response frequency ($\geq$ 100 Hz). When the duty cycle of the polarity switching is chosen to be symmetric, symmetric ion wakefields are expected to form around each dust grain (Fig.~\ref{fig:ionwake}c). The duty cycle can be tuned to cause anisotropies, which in turn modulate the dust interactions and the resulting filamentary structure \cite{dietz_phase_2021}. An analogy with liquid crystals is recognized by considering a dust particle filament (consisting of multiple dust grains) as an equivalent of an LC rod-shaped molecule. However, the mechanism described here differs from liquid crystal properties reported in complex plasma experiments with cylindrical dust particles \cite{molotkov_liquid_2000}, \cite{khrapak_liquid_1999}. 

An important distinction between PK-4 dusty plasma and conventional LCs is that an external electric field is needed to establish the interaction potential leading to the formation of filaments. This may have led previous investigations of the filamentary dusty plasma state to focus on ER fluids instead of LCs. Here we point out that the electric field in PK-4 can be thought of as the mechanism necessary to establish ``bonds'' among individual dust grains, analogous to the chemical bonds holding atoms in a molecular structure. However, the electric field alone is insufficient to cause strong alignment of the filaments along the director axis, the layering of filaments in nested structures, and the formation of hexagonal patterns in the cross-field plane. Neutral gas pressure and sufficient dust number density, in combination with changes to the interaction potential caused by the plasma dynamics, are needed to induce these more complex structural transitions.

Here we investigate the simultaneous liquid and crystalline properties of dusty plasma clouds using video data from the PK-4 facility on board the ISS \cite{pustylnik_plasmakristall-4_2016}. Unlike molecular liquid crystals, complex plasma is optically thin and macroscopic, which allows for the real-time observation of particle dynamics at the kinetic level. Particle tracking is used to obtain dust positions as a function of time for each set of experimental data. Then, the structural properties of the dust clouds are assessed by calculating pair correlation functions for particles (i) within individual filaments, (ii) within the central cloud plane, and (iii) within successive planes to reconstruct the 3D cloud behavior. The computed coupling of dust particles within filaments, across filaments in a plane, and across the bulk structure are then used to identify properties reminiscent of those observed in the nematic and smectic phases of conventional LCs. We report on both transition to a nematic and a smectic state for certain combinations of experimental parameters. These observations are confirmed by simulations of the ions and dust using the DRIAD (Dynamic Response of Ions And Dust) code. 

This paper is organized as follows. Section II provides details of the experimental setup, major observations from the experiment, and methods of data analysis using different typed of pair correlation functions. Section III introduces the simulations used to study the PK-4 discharge conditions and the dust ordering in filamentary structures. In Sec. IV, we compare experimental and simulation findings and discuss the mechanisms leading to the observed structural states. Section V provides a summary of the main findings. 

\section{\label{sec:pk4}Dust Structures in the PK-4 Experiment}
\subsection{Experimental setup}
\begin{figure}[hbt!]
\centering
\includegraphics[width=0.4\textwidth]{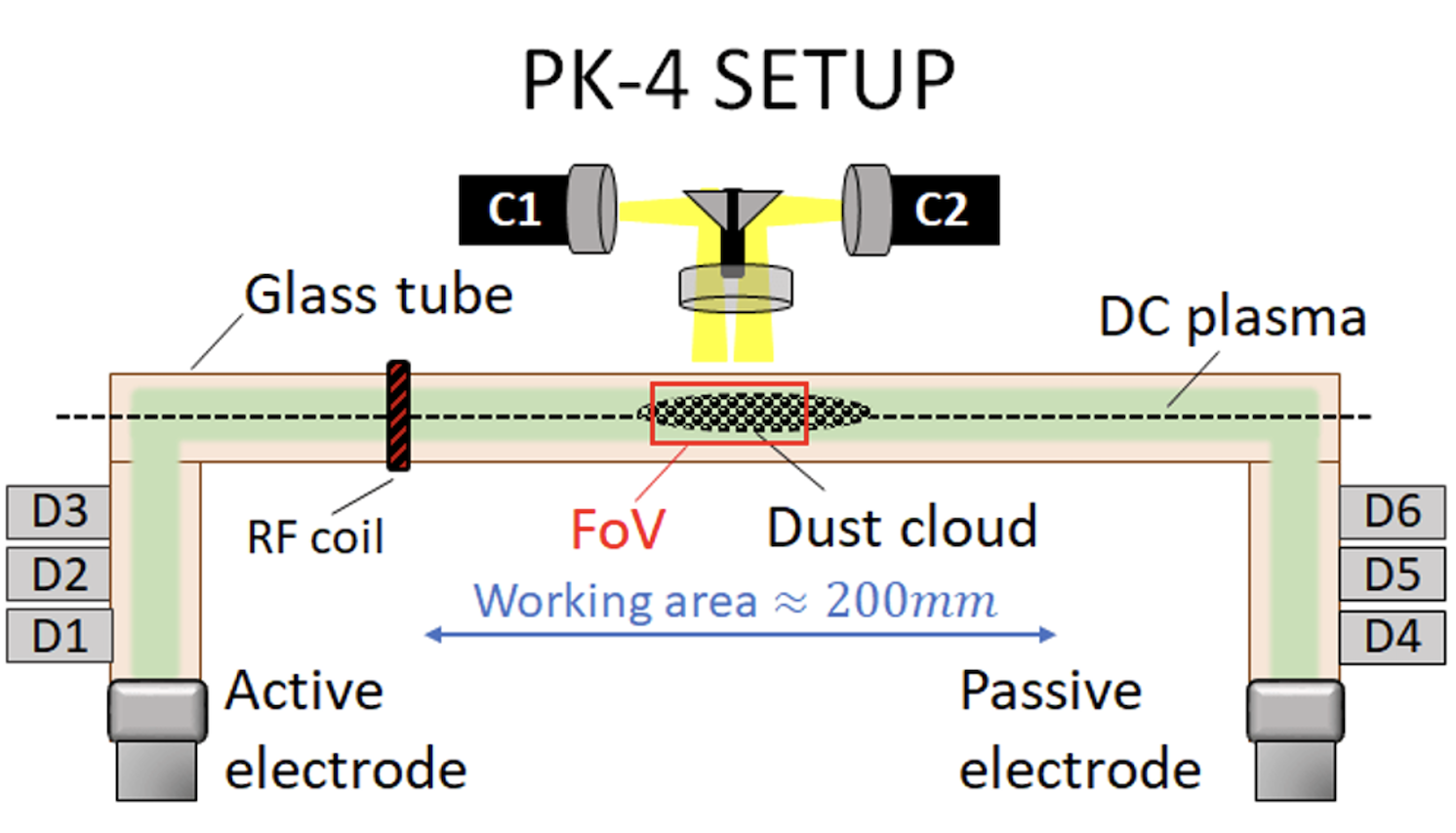}
\caption{The PK-4 setup: In the present experiments, DC plasma was created in a $\Pi$-shaped glass chamber. Dust particles were injected from Dispenser 5 (D5), one of six available dispensers. Laser illumination (not shown) and video cameras were used to observe the dust cloud in the field of view (FoV). }
\label{fig:pk4}
\end{figure}

The PK-4 laboratory is a successor of the PKE-Nefedov \cite{nefedov_pke-nefedov_2003} and PK-3 Plus \cite{thomas_complex_2008} experiments on board the International Space Station. The PK-4 experimental setup consists of an elongated $\Pi$-shaped glass chamber (85 cm length and 3 cm inner diameter) in which plasma can be ignited either by cylindrical DC electrodes or by RF coils (Fig.~\ref{fig:pk4}). The present work focuses on dusty plasma experiments utilizing pure DC neon gas discharges. A bipolar, high-voltage power supply powers the DC electrodes and provides output current as high as 3.1~mA and a maximum voltage of 2.7~kV. The DC current can either be unidirectional or polarity-switched, used to transport dust after injection or to trap/manipulate dust, respectively. In the present experiments, 500~Hz polarity switching of the DC electric field was used to trap the dust particles in the cameras' field of view. The frequency response of micron-sized dust particles is a function of the charge-to-mass ratio, usually in the range of 10-100 Hz \cite{nitter_levitation_1996}, \cite{fortov_complex_2005}. As a result, the dust particles have a negligible response to the 500~Hz polarity switching frequency and respond only to the time-averaged electric field. If an asymmetric duty cycle is used, the dust cloud experiences a net electric force and is moved along the axis of the glass chamber. With a symmetric duty cycle (50$\%$) used in the present experiments, the time-averaged electric field is zero, and the dust is trapped within the cameras' field of view.  However, since the plasma species can respond to a 500~Hz frequency, an indirect dust response to the polarity switching could be observed due to the plasma-dust interaction.  The dust particles were illuminated by a laser sheet (532~nm diode laser) and video data was collected with two identical particle observation video cameras (1600 $\times$ 1200-pixel CCD chips).

All experiments analyzed in this paper were recorded at a frame rate of 71.4~frames per second (fps) for each of the two particle observation cameras, with a combined field of view $\approx~44\times 7~\mathrm{mm}^2$ (assuming resolution of $14\times 14~\mathrm{\mu m}^2/\mathrm{px}$). In addition, the characteristics of the plasma discharge were monitored by a plasma glow observation camera ($640\times 480-$pixel CCD chip), which covers the entire working area of the experiment with a resolution of  $\approx 430 \times 430~\mathrm{\mu m}^2/\mathrm{px}$. The frame rate for the images obtained with the plasma glow observation camera is 15.2~fps. More details on the PK-4 facility can be found in \cite{pustylnik_plasmakristall-4_2016}.

\subsection{Experimental description and observed structure}
The experiments described in this paper were conducted in July, 2019, as part of PK-4 ISS Campaign 7. The same dataset has been recently used by McCabe et al. \cite{mccabe2025experiments} to investigate the dust heating due to changes in polarity switching and by Andrew et al. \cite{andrew2024anisotropic} to study anisotropic anomalous diffusion. Thus, the findings of the present paper are complimentary and directly comparable to the findings from McCabe et al. and Andrew et al. 

We focus on experiments in neon DC plasma at pressures approximately equal to 28.5~Pa, 46.1~Pa, and 70.5~Pa. For each pressure case, three values of DC current were examined, 1~mA, 0.7~mA, and 0.35~mA, yielding nine sets of distinct pressure-current conditions. The dust particles introduced in the plasma were melamine-formaldehyde spheres of diameter $3.38 \pm 0.07 \mu$m. For each pressure case, the micro-particles were injected and transported into the cameras' field of view (FoV) by a unidirectional DC field. The dust particles were then trapped in the FoV using polarity switching of the DC current with 500~Hz frequency and 50$\%$ duty cycle.  A central plane within the dust cloud was illuminated by a laser sheet. Here, the plane of the laser sheet is chosen to coincide with the x-z plane, with the y-direction perpendicular to the laser sheet (see Fig.~\ref{fig:pk4} for coordinate system). In each experiment, after the dust cloud had settled for $\sim$50~s, a Y-scan was performed by moving the laser illumination sheet in the y-direction with a scan velocity of 1~mm/s. The Y-scan procedure allowed for approximate reconstruction of the three-dimensional structure for each pressure-current case. Table~\ref{tab:exp_params} lists the pressure, current, and calculated dust density for each examined experimental data set.

\begin{table}
    \centering
    \begin{tabular}{|c|c|c|c|c|c|c|c|c|c|} 
    \hline  
         Set &  1 &  2 &  3 &  4 &  5 &  6 &  7 &  8 &  9 \\ 
    \hline  
         $P$ [Pa] &  28.5 &  28.5 &  28.5 &  46.1 &  46.1 &  46.1 &  70.5 &  70.5 &  70.5 \\ 
    \hline  
         $I$ [mA] &  0.35 &  0.7 &  1 &  0.35 &  0.7 &  1 &  0.35 &  0.7 &  1 \\ 
    \hline  
         $n$ [mm$^{-3}$] &  81.8 &  88 &  85.3 &  123.6 &  93.4 &  93.3 &  55.1 &  93.3 &  69.3 \\ 
    \hline 
    \end{tabular}
    \caption{Pressure, current, and dust density for each examined data set from PK-4 experiments.}
    \label{tab:exp_params}
\end{table}

\begin{figure}[b]
\centering
\includegraphics[width=0.45\textwidth]{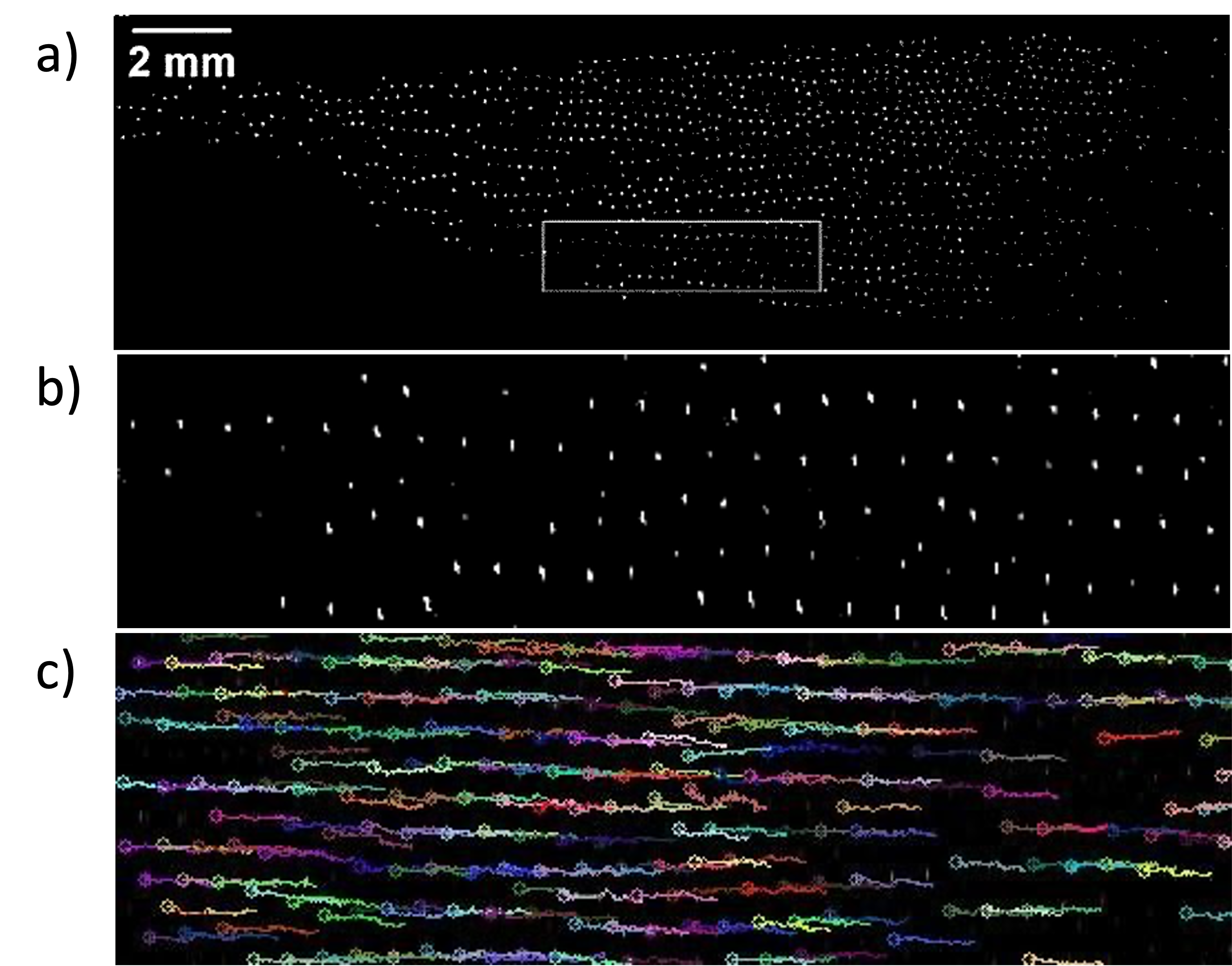}
\caption{a) Filamentary dusty plasma cloud in neon DC discharge with polarity-switched electric field. b) Zoomed in section of the cloud showing filamentary structure along the director axis. c) Trajectories of particles within the cloud over $\approx$ 1.5 s period. Data from PK-4 experiment with pressure 70.5 Pa and DC current 1 mA.}
\label{fig:dusttrack}
\end{figure}

In all cases, the formation of filamentary structures in the xz-plane was observed upon the application of the symmetric polarity-switched DC current. Similar to ER fluids, the transition to a filamentary state in PK-4 dusty plasmas is enhanced by the induced electric field, which alternates directions along the x-axis as the DC polarity is switched. In addition, similar to LCs with rod-shaped molecules in the nematic phase, the dust filaments in PK-4 exhibited a common orientation along a director axis (here, the x-axis) which coincides with the direction of the externally applied electric field. Fig.~\ref{fig:dusttrack}a shows an image of a typical filamentary dust cloud in neon plasma at pressure 70.5~Pa and DC current 1~mA. For these conditions, the extended filaments consisted of 20-30 particles each (Fig.~\ref{fig:dusttrack}b). Fig.~\ref{fig:dusttrack}c shows the particle trajectories over a period of $\sim 1.5$~seconds (camera frame rate 71.4~fps) in the small region of the cloud highlighted by a white rectangle in Fig.~\ref{fig:dusttrack}a. It can be seen that the dust particles remain aligned within filaments as they move, suggesting a stable structure.  

\begin{figure}[bt!]
\centering
\includegraphics[width=0.45\textwidth]{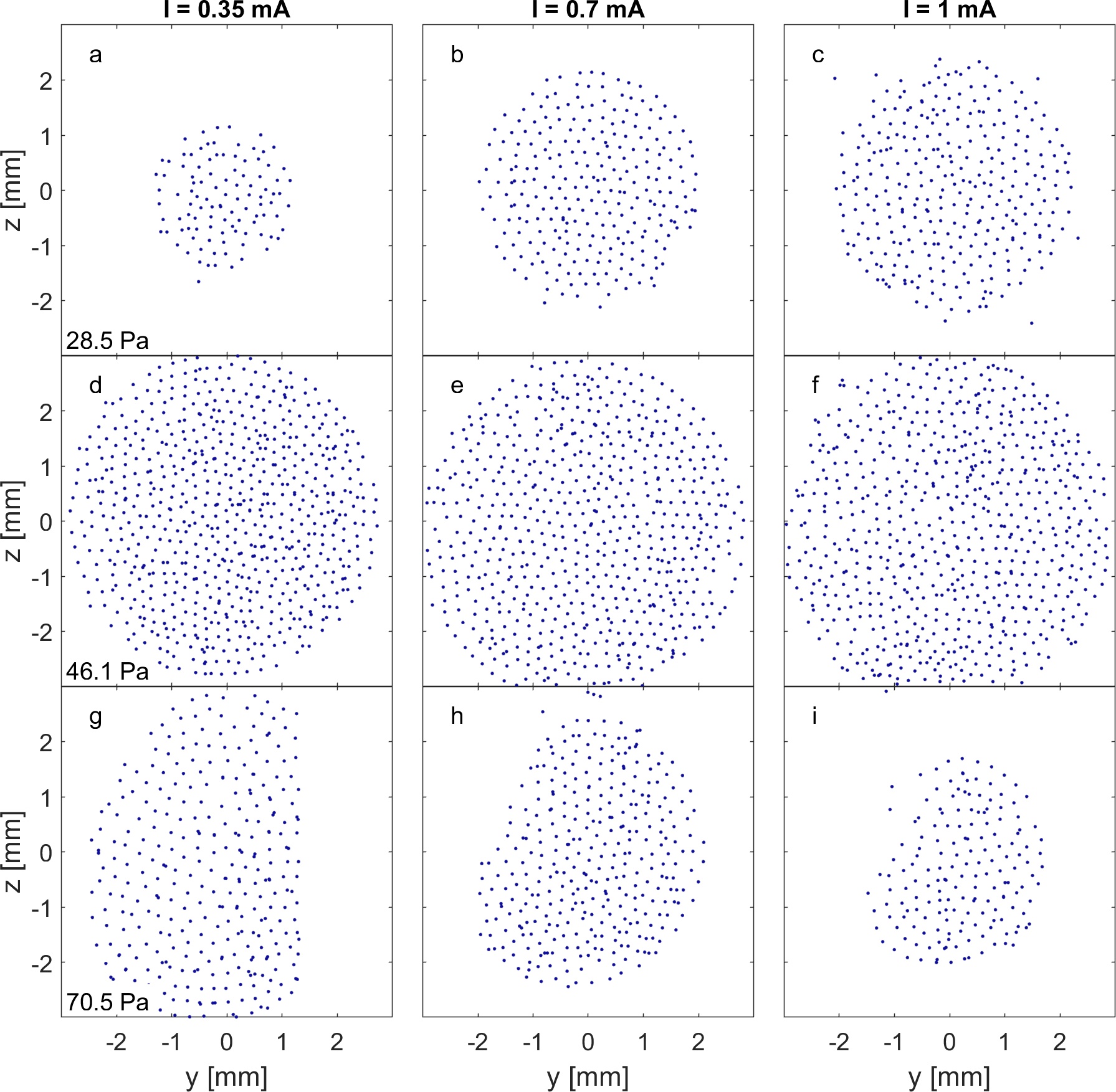}
\caption{Cross sections of the trapped dust clouds for each set of  experimental conditions. Reconstruction of the particle positions was obtained from particle tracking during a Y-scan using a $250~\mu$m-thick slice in the x-direction.}
\label{fig:dustcloud}
\end{figure}

Fig.~\ref{fig:dustcloud} shows the cloud structure in the yz-plane (perpendicular to the external electric field) for each pressure-current case. The images were reconstructed using particle positions within a $100~\mu$m-thick slice. This data was extracted for successive images obtained during Y-scans. At pressures 28.5 Pa and 46.1 Pa, the clouds had an ellipsoid shape in the xz-plane and exhibited circular symmetry in the yz-plane for all current conditions. At 70.5 Pa, the yz-cross-section of the clouds was asymmetric, which could have resulted from a plasma inhomogeneity.  Filamentary structures were observed to form in the xz-plane for all pressure-current cases except for 28.5 Pa, 0.35~mA where the cloud was the smallest and distribution of particles was mostly isotropic. At pressures of 28.5 Pa and 46.1 Pa, the diameter of the clouds in the yz-plane increased with the increase of the DC current (Fig.~\ref{fig:dustcloud}a-f), while at 70.5 Pa, the diameter of the clouds decreased with current (Fig.~\ref{fig:dustcloud}g-i). While there are always differences in the dust injection procedure from case to case, the resulting cloud size is largely affected by the structure of the discharge glow. In the 28.5 Pa and 70.5 Pa cases, the discharge exhibited large-scale stratification resulting in multiple smaller dust clouds located in neighboring striations. In contrast, the discharge was mostly homogeneous in the 46.1 Pa case yielding one large cloud. For this pressure, the clouds also exhibited higher dust number density, as can be seen from Table~\ref{tab:exp_params} (highest dust density for the 0.35 mA case).

\begin{figure}[hbt!]
\centering
\includegraphics[width=0.45\textwidth]{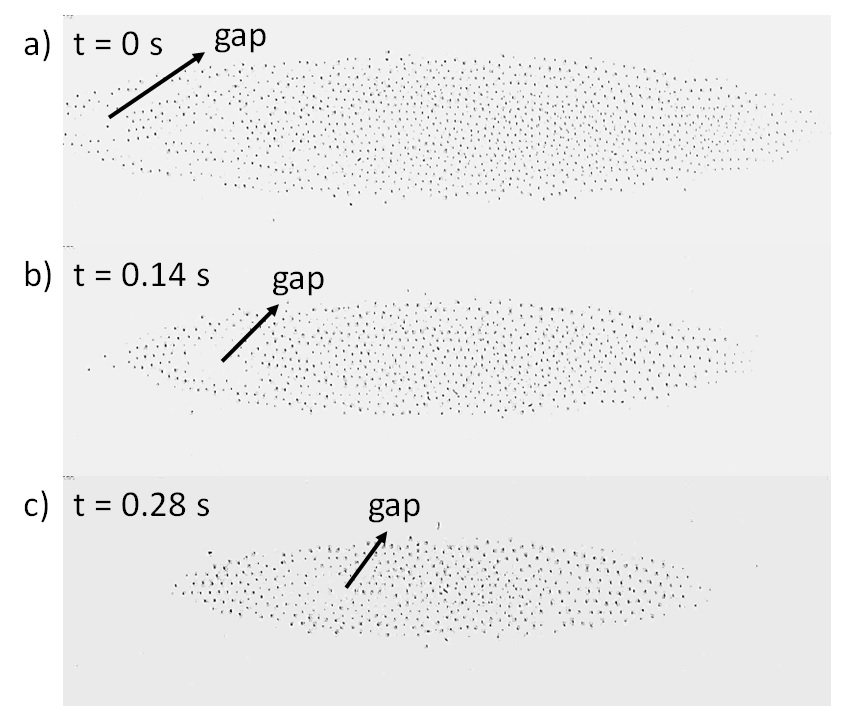}
\caption{Successive images of the dust cloud taken 0.14 s apart during a Y-scan with the particle observation laser sheet (laser current 0.7 mA and scan velocity 1 mm/s). Data is presented for 1 mA DC current and pressure 46.1 Pa. The approximate laser offset from its initial position at the center of the discharge tube is a) 2.68 mm, b) 2.82 mm, and c) 2.96 mm. The images were recorded with a single particle observation camera with a field of view $\approx$ 22.6 $\times$ 6.8 mm$^2$. As time increases and the laser sheet moves away from the central axis, clear gaps (void of particles) are observed on the left side of the cloud.}
\label{fig:cloud_gaps}
\end{figure}

In Fig.~\ref{fig:dustcloud}, ring structures can be observed in the exterior of the clouds suggesting that the clouds consist of nested surfaces. The presence of large-scale layers, or shells (smectic behavior) is also observed from video data during the Y-scan. Fig.~\ref{fig:cloud_gaps} shows successive video frames obtained as the laser sheet illumination moves from the center to the exterior of the cloud during the Y-scan procedure. These images reveal the presence of a particle-free voids within the 3D structure, suggesting that the dust cloud consists of several nested spheroids.

\subsection{Pair correlation analysis of experimental data}

 Two types of video data were analyzed to characterize the structure of the clouds: (i) videos where the laser sheet was fixed in the central region of the cloud and (ii) videos from Y-scan procedures, where the cameras and laser sheet were moved across the cloud in the y-direction, which allows for the study of the three-dimensional structure. The particle positions in the x-z plane were obtained using the particle tracking MosaicSuite plugin of Fiji, which is a distribution of ImageJ \cite{mosaic_fiji_2013}. The dust positions in $\mu$m were obtained assuming a pixel resolution of $14.20~\mathrm{\mu m}$ \cite{pustylnik_plasmakristall-4_2016}. The dust positions in the y-direction were determined by the location of the laser sheet. Given that the laser moved with a constant speed of 1.0~mm/s and that the video frame rate was 71.4~fps, we assumed a displacement of about $14~\mathrm{\mu m}$ between frames. As the laser sheet scans through the cloud, some particles may be picked up in multiple, consecutive frames. To ensure that the analysis does not double count particles, we filtered out particles that are less than a threshold distance from another particle in the next few subsequent frames. This process yields a three-dimensional reconstruction of the dust cloud (Fig.~\ref{fig:dustcloud}).  The 3D dust positions are used to perform structural analysis by calculating pair correlation functions.     

The pair correlation function represents the probability of finding a neighbor particle at a distance $r$ away from any given particle in the cloud, thus providing information about the structural arrangement. A completely random distribution of particles, like the one for a gaseous state, has a pair correlation function that looks like a fluctuating flat line, indicating uncorrelated positions as a function of space. For a periodic crystal, the pair correlation function appears as a series of periodic, well-separated sharp peaks with troughs dropping to zero in between peaks. For a liquid-like structure, the peaks are less pronounced, and the troughs do not drop to zero. This indicates that the particles are correlated, but also have a considerable degree of freedom to move in space. 

The radial pair correlation function for particles in a 2D plane, $g(r)$ is defined as
\begin{equation}
    g(r) = \frac{1}{n_d N} \sum_{i \neq j}^N \frac{\delta(r-r_{ij})}{2\pi r \Delta r},
\label{eq:gr_1D}
\end{equation}
where $n_d$ represents the 2D number density, $N$ is the total number of particles, and $r_{ij}$ is the distance between particles $i$ and $j$. The normalization reflects the average number of particles expected to occupy an annular ring centered on particle $i$ with radius $r$ and width $\Delta r$. 

\begin{figure}[t]
\centering
\includegraphics[width=0.45\textwidth]{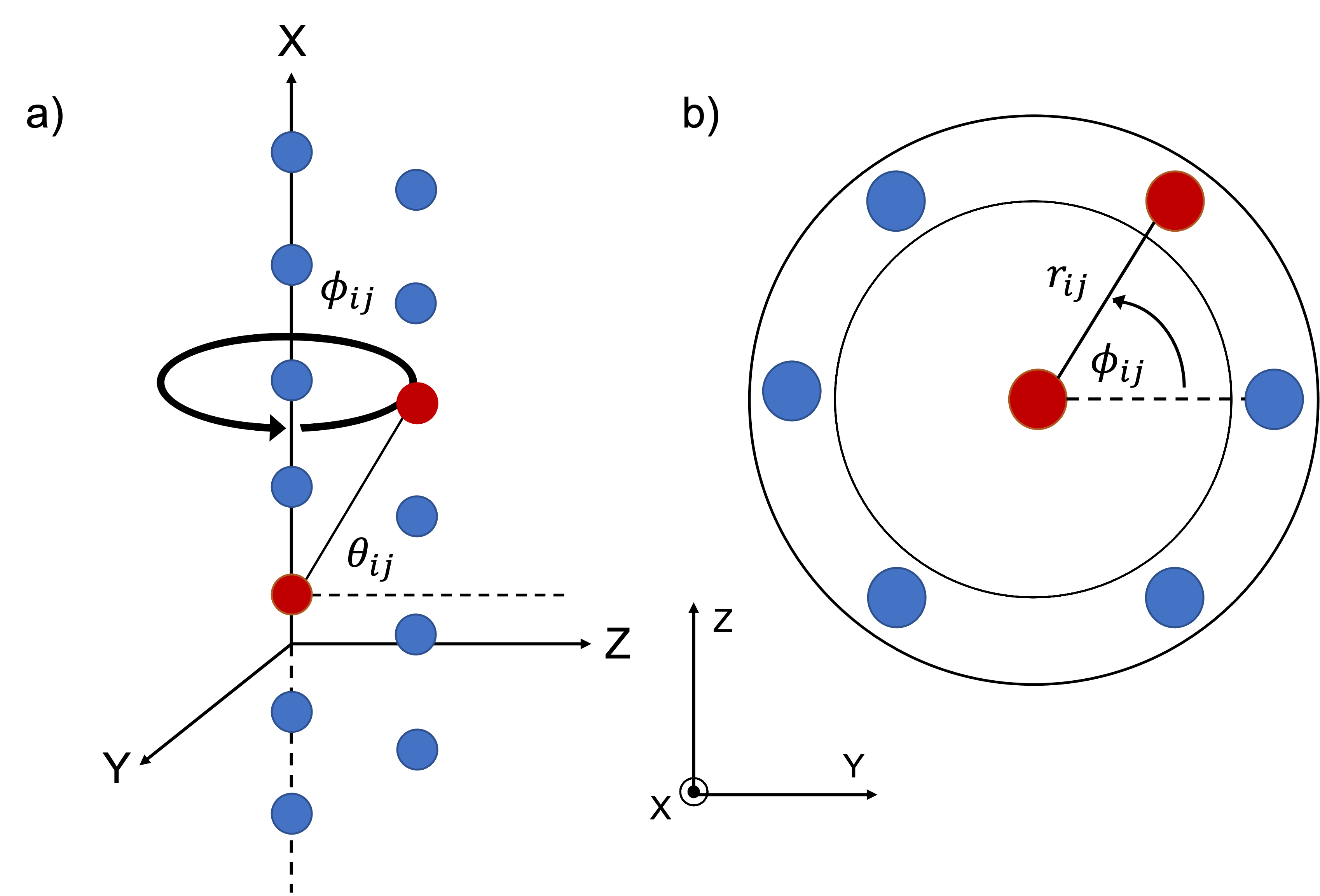}
\caption{Coordinate system used for three-dimensional pair correlation functions illustrated by a) a side view, with filaments aligned with the x-axis, and b) a cross section of the cloud with the dust filaments extending perpendicular to the page.}
\label{fig:coords}
\end{figure}

A 3D pair correlation function in spherical coordinates was introduced in \cite{pustylnik_three-dimensional_2020} and \cite{mitic_long-term_2021}

\begin{equation}
    G_\phi(r,\theta) = \frac{1}{n_dN} \sum_{i,j=1, i \neq j}^N \frac{\delta(r_{ij} - r) \delta(\theta_{ij} - \theta)}{2\pi r^2 \cos(\theta) \Delta r \Delta \theta}
\label{eq:Gphi}
\end{equation}

\begin{equation}
    G_\theta(r,\phi) = \frac{1}{n_dN} \sum_{i,j=1, i \neq j}^N \frac{\delta(r_{ij} - r) \delta(\phi_{ij} - \phi)}{2 r^2 \Delta r \Delta \phi},
\label{eq:Gtheta}
\end{equation}
where $n_d$ is the 3D number density, $N$ is the total number of dust particles, and $\Delta r$, $\Delta \theta$, $\Delta \phi$ represent the bin width for $r$, $\theta$, $\phi$, respectively. The polar angle $\theta$ ranges from $\{-\frac{\pi}{2}, \frac{\pi}{2}\}$ radians, with $\theta = 0$ on the yz-plane centered on the $i^{th}$ particle (Fig.~\ref{fig:coords}a). The polar angle can be used to determine the strength and range of coupling among dust particles within a filament (at the angles $\theta \approx \pm \pi/2$), as well as particles in neighboring filaments. The azimuthal angle $\phi$ ranges from $\{0, 2\pi\}$ radians as measured from the y-axis in the yz-plane (Fig.~\ref{fig:coords}b). The azimuthal angle provides information about the relative orientation of neighboring filaments, revealing characteristics of the layered structure and hexagonal symmetry (if present).

The radial pair correlation function $G(r)$ for particles in a 3D volume is calculated by 
\begin{equation}
    G(r) = \frac{1}{n_dN} \sum_{i,j=1, i \neq j}^N \frac{\delta(r_{ij} - r) \delta(\phi_{ij} - \phi)}{4 r^2 \Delta r},
\label{eq:G_r}
\end{equation}
which differs from the 2D pair correlation function $g(r)$ in that the data is normalized by the number of particles expected to occupy a spherical shell with radius $r$ and width $ \Delta r$ centered on particle $i$.

\subsubsection{Radial Pair Correlation}
The average interparticle separation can be determined from the location of the first peak of $g(r)$. As $g(r)$ is calculated for particles in successive planes across the cloud, shifts in the location of this peak can be used to identify gradients in dust density and the existence of layers. 

\begin{figure}[hbt!]
\centering
\includegraphics[width=0.45\textwidth]{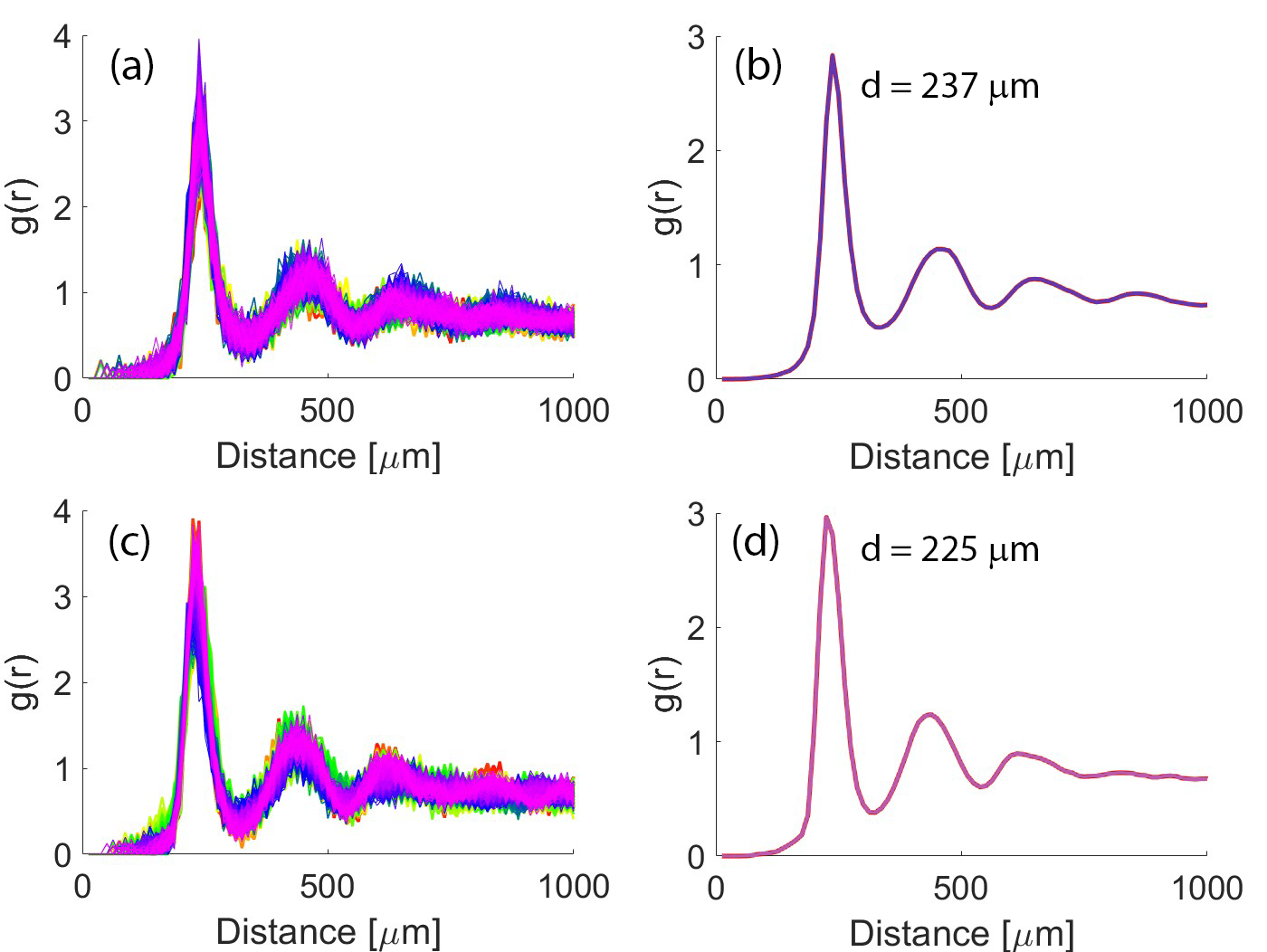}
\caption{2D radial pair correlation functions $g(r)$ for particles within a fixed 2D central plane at a),b) 28.5 Pa, 0.7 mA and c),d) 70.5 Pa, 0.7 mA. The colors in a) and c) indicate the progression of time. The corresponding time-averaged pair correlation functions are shown in b) and d). The location of the first peak in b) and d) is used to determine the average interparticle spacing in the cloud, as indicated on the plots.}
\label{fig:avgintsep}
\end{figure}

Data from dust particles within a central plane of the dust cloud were analyzed for long periods (20 s, 71.4 fps) when no Y-scan was taking place. In these data sets, most of the dust particles remained in the field of view for the entire image sequence.  Fig.~\ref{fig:avgintsep} shows representative results for $g(r)$ calculated for the 28.5~Pa and 70.5~Pa cases, both at 0.7 mA. Fig.~\ref{fig:avgintsep}a,c show the stacked $g(r)$ for individual frames where the color change indicates the progression in time. It can be seen that the pair correlations do not vary appreciably over extended periods of time for both pressure cases. Similar trends are observed for all pressure-current data sets. Fig.~\ref{fig:avgintsep}b,d show the time-averaged pair correlations and average interparticle separation $d$ determined from the location of the first peak. It can be seen that $d$ decreases slightly as the pressure is increased, likely due to a change in the particle number density. There are four distinct peaks at the lower pressure (Fig.~\ref{fig:avgintsep}b), while the pair correlation at high pressure (Fig.~\ref{fig:avgintsep}d) has only three clearly distinguishable peaks, suggesting a greater range of the interparticle coupling withing the xz-plane for the lower pressure case. This is a result of the $g(r)$ being obtained as averages over all particles within the xz-plane. In the lower pressure case, the observed structure was more isotropic with similar coupling of particles within and across filaments. At high pressures, the filamentary structures were more pronounced resulting in stronger coupling within filaments and weaker coupling across filaments. In the next section, we show pair correlations in spherical coordinates, which support these observations.

\begin{figure}[hbt!]
\centering
\includegraphics[width=0.45\textwidth]{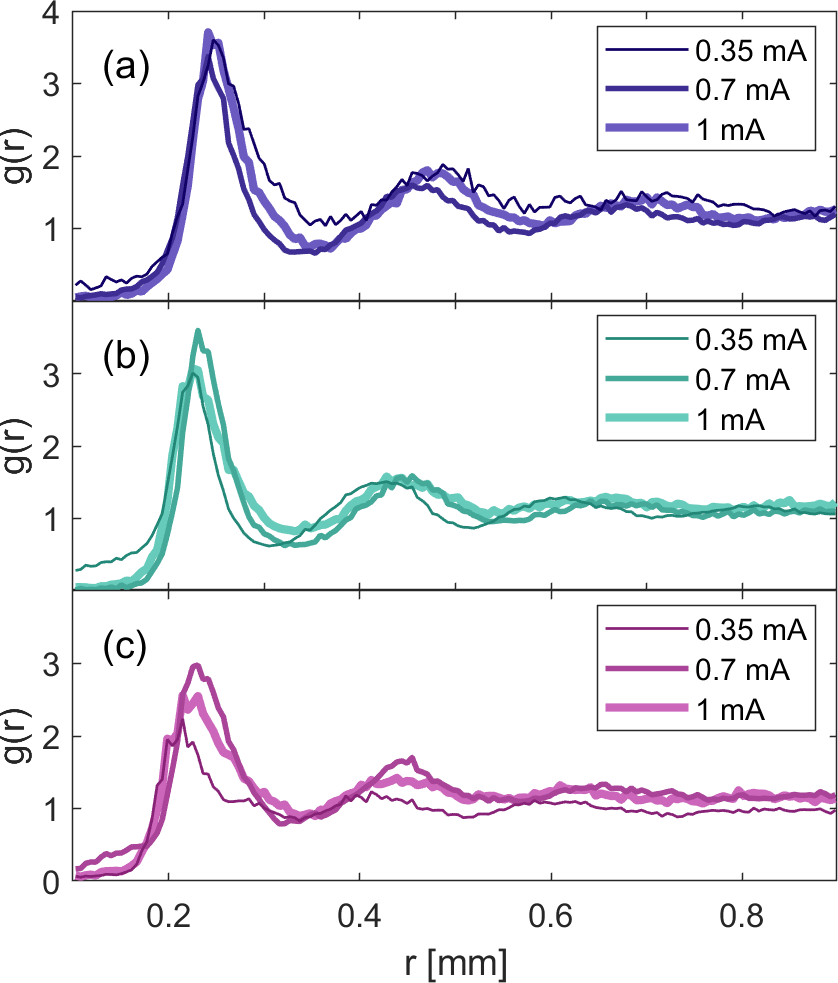}
\caption{Two-dimensional radial pair correlation functions $g(r)$ averaged over 29 frames from a Y-scan covering a 400-$\mu$m-thick slice at the center of each cloud for the three different pressures a) 28.5 Pa, b) 46.1 Pa, and c) 70.5 Pa and all DC current values.}
\label{fig:2Dgr_Yscan}
\end{figure}

The small variation in time shown in Fig.~\ref{fig:avgintsep} validates the assumption that the cloud is approximately stationary over the characteristic time scale of a Y-scan (several seconds). This allows for a meaningful comparison of $g(r)$ values calculated for particles in successive 2D planes using data from the Y-scans. Fig.~\ref{fig:2Dgr_Yscan} shows the time-averaged 2D g(r) plots calculated from 29 successive frames covering a 400-$\mu$m-thick central section of the cloud during a Y-scan. 
At 28.5 Pa (Fig.~\ref{fig:2Dgr_Yscan}~a), the interparticle spacing does not change appreciably with discharge current, which can be explained by the overall similar dust number density in each case (see Table~\ref{tab:exp_params}). At 28.5 Pa, the peaks in $g(r)$ are more pronounced at higher current, suggesting enhancement of the long-range order in the structure. At 46.1 Pa (Fig.~\ref{fig:2Dgr_Yscan}~b), the interparticle separation does not appreciably change with current. However, at this pressure, the most pronounced peaks (and corresponding long-range order) are observed for the 0.35 mA case, which is likely due to the much higher dust density in this case. At 70.5 Pa (Fig.~\ref{fig:2Dgr_Yscan}~c), there is a more pronounced change of the interparticle separation with changing current due to the large difference in the dust densities. The $g(r)$ peaks at 70.5 Pa seem to be most pronounced at 0.7 mA, which is also the case with highest number density for this pressure. Thus, we conclude that the interplay between current and dust density dictates the resulting interparticle separation in each pressure case. Long-range order seems to be enhanced for higher dust number density. For similar dust densities (as in the 28.5 Pa cases), the long-range order is enhanced with increasing current.

\begin{figure}[hbt!]
\centering
\includegraphics[width=0.45\textwidth]{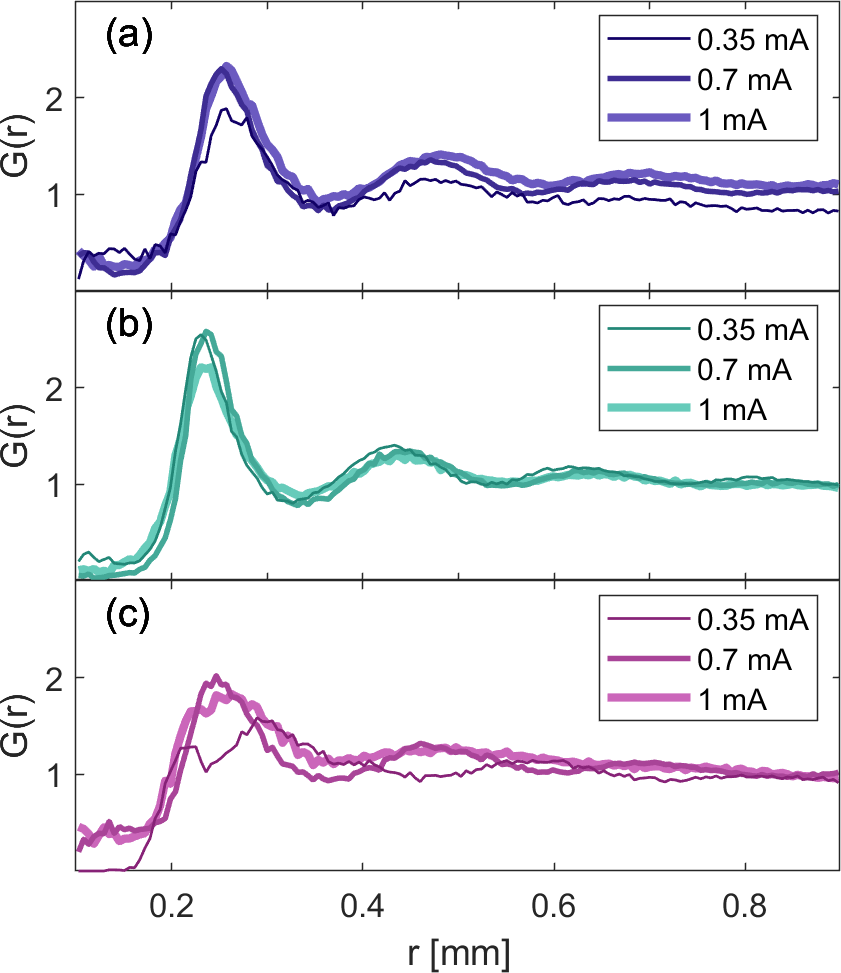}
\caption{Three-dimensional pair correlation functions $G(r)$ calculated from reconstructed 3D dust clouds at the three different pressures a) 28.5 Pa, b) 46.1 Pa, and c) 70.5 Pa and all DC current values.}
\label{fig:gr3D}
\end{figure}

The calculated $G(r)$ for the reconstructed 3D clouds are shown in Fig.~\ref{fig:gr3D}.  In contrast to $g(r)$ calculated for particles in 2D planes, the primary peak is broader, indicating that the average interparticle spacing varies from the interior to the exterior of the cloud and that there is less correlation between particle positions in neighboring filaments. In particular, at the highest pressure (Fig.~\ref{fig:gr3D}c), the second and third peaks are barely visible, indicating liquid-like behavior for the 3D cloud.  It is interesting to note that at 70.5 Pa and 0.35 mA, the primary peak splits into two closely spaced peaks. This is indicative of distinct directional anisotropy in the interparticle separation, which is further explored in the following section. 

\subsubsection{Angular Pair Correlations}

\begin{figure*}[h!tb]
\centering
\includegraphics[width=1\textwidth]{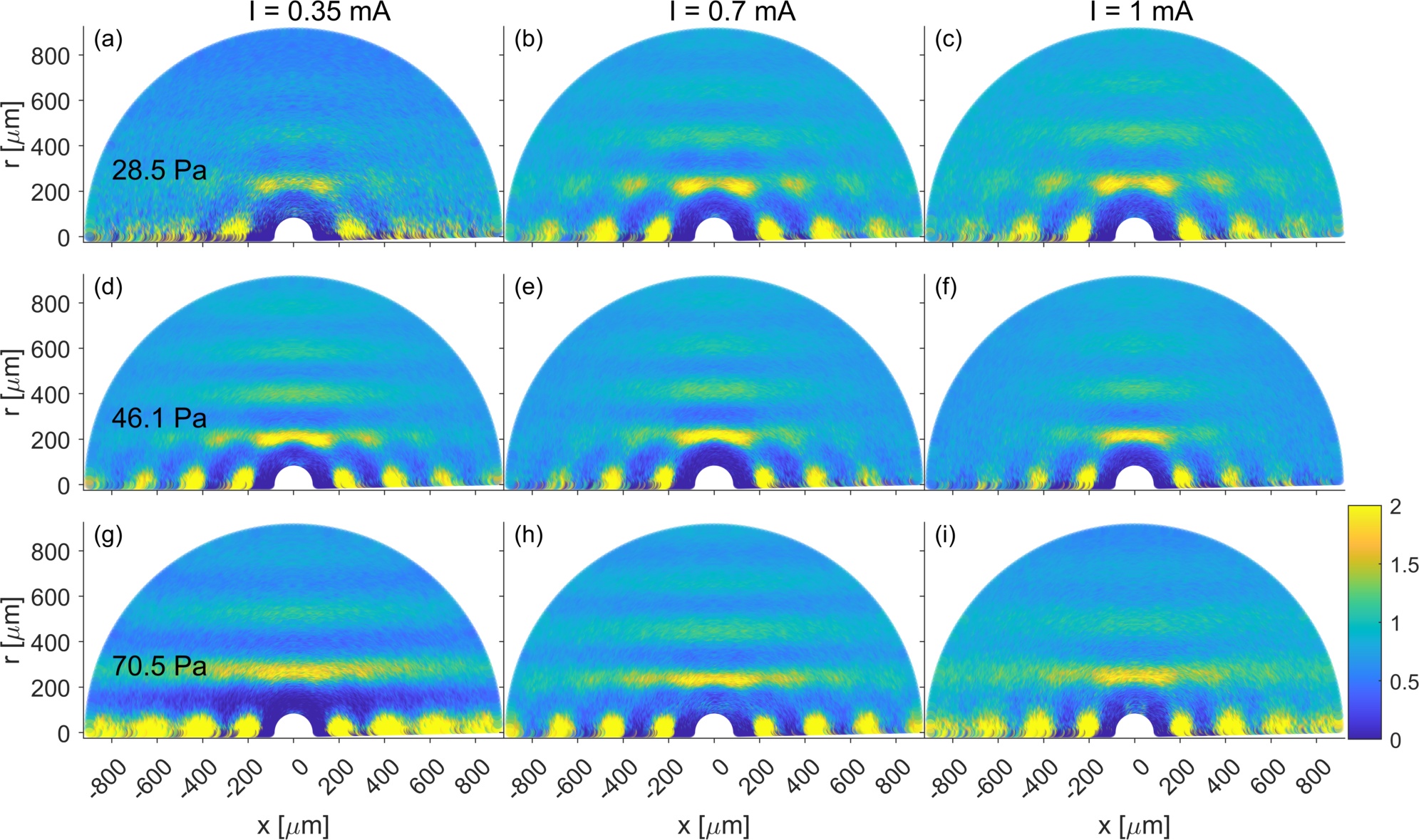}
\caption{Plots of $G_\phi(r, \theta)$ for three different pressures: a-c) 28.5 Pa, d-f) 46.1 Pa, and g-i) 70.5 Pa. Current increases from left to right. The colorbar is saturated at $G_\phi = 2$ to reveal structure at large distances.  The abscissa denotes the distance along the director axis while the ordinate denotes the radial distance away from each particle in the x-z plane. Note that the data shown in (a) is noisy due to the small number of dust particles within the cloud.}
\label{fig:gphi}
\end{figure*}

\begin{figure*}[h!tb]
\centering
\includegraphics[width=1\textwidth]{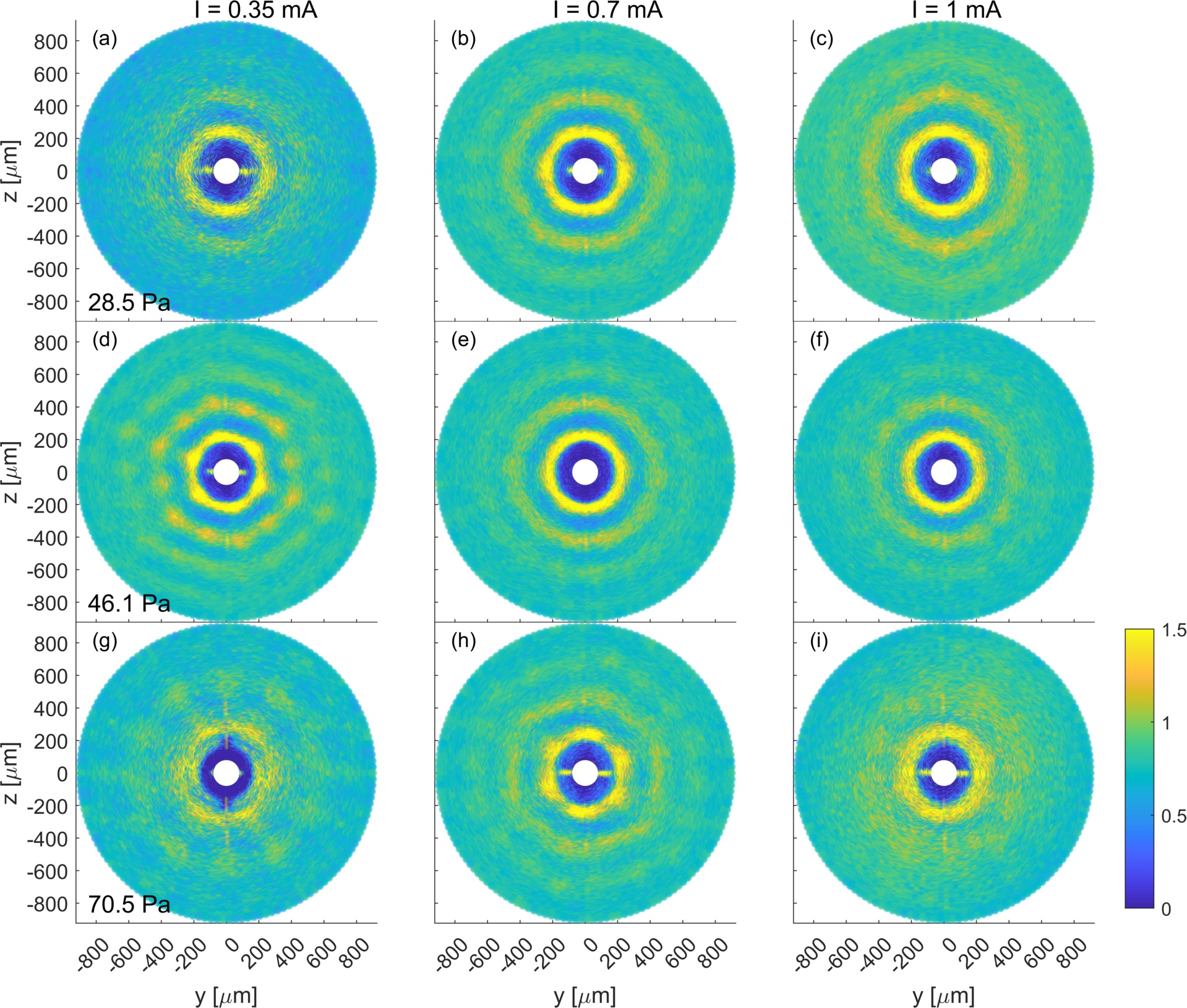}
\caption{Plots of $G_\theta(r, \phi)$ for three different pressures: a-c) 28.5 Pa, d-f) 46.1 Pa, and g-i) 70.5 Pa. Current increases from left to right. The colorbar is saturated at $G_\theta = 1.5$ to reveal structure at large distances.  Note that the data shown in (a) is noisy due to the small number of dust particles within the cloud.}
\label{fig:gtheta}
\end{figure*}

Further information about the 3D structural characteristics of the clouds is obtained from the the 3D angular pair correlation functions $G_\phi(r,\theta)$ from Eq.~(\ref{eq:Gphi}) and $G_\theta(r,\phi)$ from Eq.~(\ref{eq:Gtheta}). Fig.~\ref{fig:gphi} shows $G_\phi(r, \theta)$ and Fig.~\ref{fig:gtheta} shows $G_\theta(r, \phi)$ for all pressure-current conditions.  The data for $G_\phi(r,\theta)$ is plotted in $(x,r)$ coordinates in Fig.~\ref{fig:gphi}, where $x$ is the distance along the x-axis (along the direction of the external electric field) and $r$ is the radial distance from the x-axis (see Fig.~\ref{fig:coords}a). Particles within a given filament aligned with the x-axis will be located approximately at $\theta = 0$ and $\theta = \pi$, which coincides with $r = 0$ in Fig.~\ref{fig:gphi}. The periodic bright spots at distinct positions along the x-axis were referred to as ‘string peaks’ in \cite{mitic_long-term_2021}. The string peaks indicate periodic crystalline order within field-aligned filaments.  The location of the first string peak on either side of $x = 0$ gives the average interparticle separation within the filaments. The location of peaks at angles $0 < \theta < \pi$ ( or $r > 0$ in Fig.~\ref{fig:gphi}) shows alignment between particles within nearby filaments. 

At 28.5 Pa (Fig.~\ref{fig:gphi}~a-c), the dust clouds for all currents exhibit strong coupling between nearest neighbor particles within filaments and across filaments, which is evident from the periodic patterns of bright spots at $r = 0$ and $r \approx 220~\mu m$. The 0.35 mA case (Fig.~\ref{fig:gphi}~a) shows distinct and symmetric coupling for only the first nearest neighbors (short-range coupling). The 0.7 mA and 1 mA cases exhibit one more peak along the x-axis, suggesting the emergence of a directional anisotropy in the dust coupling, which indicates the formation of stable filamentary structures. Thus, we conclude that the structure at 28.5 Pa and 0.35 mA is a crystal with symmetric short-range coupling, while the 0.7 mA and 1 mA cases for that pressure transition to a crystal with anisotropic coupling. The transition to anisotropic coupling is likely due to a combination of increased current and dust number density. 

At 46.1 Pa (Fig.~\ref{fig:gphi}~d-f), the distinct spots at $r \approx 200~\mu m$ merge into a short band extending in the range $-200~\mu m < x < 200~\mu m$. This indicates that the coupling of particles across filaments becomes weaker and the filaments have more freedom to slide past each other. For this pressure, the string peaks along the x-axis are still pronounced, indicating strong coupling and crystalline order of dust particles within individual filaments. These trends are further enhanced at 70.5 Pa (Fig.~\ref{fig:gphi}~g-i) where the bright band at $r = 200~\mu m$ is elongated and string peaks along the x-axis are most pronounced. In addition, for this pressure, a distinct fourth string peak is visible for each current case, suggesting longer-range coupling as compared to the other two pressure cases. In other words, we observe periodic crystalline coupling within filaments and liquid-like coupling across filaments, which is analogous to LCs with rod-like molecules. 

Based on Fig.~\ref{fig:gphi}, we conclude that: (i) above a critical dust density and DC current, anisotropic coupling occurs for all pressure-current cases, which leads to filament formation and (ii) increasing pressure results in a transition from anisotropic crystalline to a liquid crystal structure. However, to determine the LC phase (e.g., nematic, smectic, etc.), the orientation and coupling of filaments in the y-z plane should be examined using $G_\theta(r, \phi)$.

For all pressure-current cases, the $G_\theta(r, \phi)$ plots in Fig.~\ref{fig:gtheta} show symmetric bright bands at periodic distances in the y-z plane. The number of bright bands can be interpreted as a range of the coupling within the plane, while the structure of the bands (continuous or spot-like) can be interpreted as crystalline-like versus liquid-like coupling. 

Most pressure-current cases show three bright bands indicative of long-range coupling. One exception is the 28.5 Pa, 0.35 mA case (Fig.~\ref{fig:gtheta}~a), where data is noisy due to the small particle number. Two other exceptions are 70.1 Pa (Fig.~\ref{fig:gtheta}~g), 0.35 mA and 70.1 Pa, 1 mA (Fig.~\ref{fig:gtheta}~i) which have only two bands, both of which more diffuse than in the other cases. In addition, the second outer band for these cases seems spot-like (more pronounced in Fig.~\ref{fig:gtheta}~g), suggesting short-range crystalline-like coupling. These two cases exhibit the lowest dust number density, as can be seen in Table~\ref{tab:exp_params}.

Several combinations of pressure-current conditions yield a a clear six-fold-symmetry of particle locations within the y-z plane. The most prominent example is the 46.1 Pa, 0.35 mA case (Fig.~\ref{fig:gtheta}~g), where the six-fold order seems to be long-range (visible as spots in the three bands). This suggests a hexagonal alignment of the filaments in the plane perpendicular to the director axis, which is reminiscent of the smectic phase in LCs. While there is not a clear trend with either the pressure or current, the long-range hexagonal symmetry is most pronounced in the case with the highest particle number density. Finally, it is curious to notice that the orientation of the six-fold symmetry varies across conditions (compare Fig.~\ref{fig:gtheta}~b, c, d, and h), which hints towards a possibility to achieve different smectic phases in dusty plasma by varying the plasma conditions and the dust density.

Based on Fig.~\ref{fig:gtheta}, we conclude that: (i) above a critical dust density, long-range order in the y-z plane (cross-field direction) is observed for all pressure-current cases and (ii) some combinations of pressure, current, and dust density conditions lead to the formation of a smectic-like structural state with six-fold symmetry. For additional insight into the behavior of the dust cloud, we turn to the results of numerical modeling. 

\clearpage

\section{Simulating Dusty Plasma in the PK-4 Experiment}
	
	\subsection{Simulating the PK-4 Discharge}
	The positive column of the plasma generated in the PK-4 experiment appears to be close-to-homogeneous (or exhibiting low frequency standing striations) at the timescales observed by the onboard imaging equipment. However, particle-in-cell with Monte Carlo collisions (PIC/MCC) simulations of the plasma conditions present in the PK-4 have revealed the presence of fast-moving inhomogeneous features, or ``ionization waves'' \cite{hartmann_ionization_2020}. High-speed video data obtained using a ground-based analogue of the PK-4 flight version, the PK-4 BU \cite{schmidt_discharge_2020}, confirmed the presence of ionization waves.
    
	
	
\begin{figure}[hbt!]
		\centering
		\includegraphics[width=0.47\textwidth]{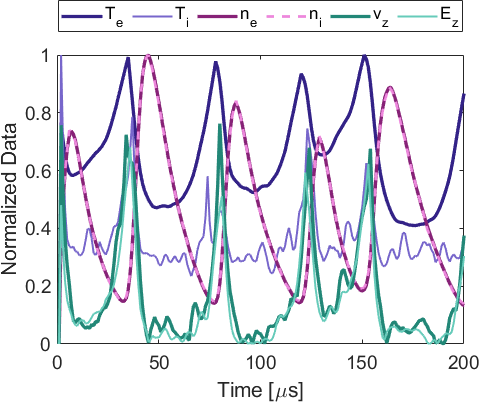}
		\caption{Results from the PIC/MCC simulation of PK-4 conditions at 60 Pa, 2 mA showing the variation in ion and electron temperature ($T_i$ and $T_e$), ion and electron density ($n_i$ and $n_e$), axial electric field ($E_x$), and ion flow velocity ($v_x$) as ionization waves pass through the plasma. All quantities are normalized by their maximum value.}
		\label{fig:ionwaveparam}
	\end{figure}

	The ionization waves move through the positive column with phase velocities in the range $500-1,200$~m/s and cause large variations in the electric field strength and plasma density \cite{hartmann_ionization_2020}. Fig.~\ref{fig:ionwaveparam} shows PIC/MCC results for a representative time series of the electric field strength, electron and ion number density, electron and ion temperature and ion flow velocity, with all values normalized by the maximum value attained during the time interval.   
	
	\subsection{Simulation of Dust and Ion Interactions}
	
	Dust grains immersed in the PK-4 experiment can respond to changes in the plasma environment on a timescale of a few milliseconds. Therefore, a common assumption is that the plasma can be reasonably modeled using the time average of the local plasma density and temperature. This assumption was shown to be inconsistent with results obtained from simulations of dusty plasmas that investigated the effects of the ionization waves on dust structure formation in PK-4 \cite{vermillion_influence_2022}, \cite{matthews_effect_2021}.
	
	The ionization waves lead to dynamically evolving dust charge and ion wakes. The fluctuating dust charge influences the charge gradient and spatial extent of the ion wake, while the ion wake mediates interactions among neighboring dust grains. To capture these effects, the evolving plasma conditions from the PIC/MCC PK-4 discharge simulation were used as input parameters in an N-body simulation of the dust and ions that models a small cylindrical section at the center of the discharge.
	
	The code Dynamic Response of Ions and Dust (DRIAD) is an N-body simulation designed to model the dusty plasma environment. DRIAD simulates the motions of individual dust grains and ions on their own timescales, as well as the dynamic charging of dust grains \cite{matthews_dust_2020}. In order to focus on the relevant time and spatial scales for dust and ion dynamics, the electrons are treated as a fluid governed by a Boltzmann distribution, and not directly modeled by the simulation. As ions and dust grains move throughout the simulation region, the forces acting on each particle are calculated in response to the spatially and temporally evolving ion number density in the ion wakefield and evolving dust grain charge. At the same time, the dust grain charge is dynamically updated based on the sum of the ion and electron currents to the dust grain. The ion current is directly obtained by tracking the number of ions that cross the collection radius of a dust grain, while the electron current is calculated using orbital motion limited (OML) theory \cite{vermillion_influence_2022}, \cite{matthews_dust_2020}, \cite{matthews_effect_2021}.
	
	The motions of ions and dust grains occur on timescales that differ by many orders of magnitude ($\Delta t_i \approx 0.01~\mu$s and $\Delta t_d \approx 0.1$ ms), which can make it difficult to model both species with a single simulation. The method used by DRIAD follows a two-step process. First, ions move and reach an equilibrium distribution over $100-1000 \Delta t_i$ in the presence of stationary dust grains. Next, the ions' positions are fixed while the dust grains move for one $\Delta t_d$ in the new ion distribution. The iterative process of exclusively moving the ions followed by exclusively moving the dust grains continues until the dust and ions have reached equilibrium. The number of dust time steps required to reach a static or dynamic equilibrium varies based on the specific conditions being simulated. 
	
	DRIAD is implemented using an asymmetric molecular dynamics scheme following the method described by Piel \cite{piel_molecular_2017}. The force on dust grains due to ions is calculated assuming a shielded Coulomb interaction, while the force on ions due to dust grains is calculated using a bare Coulomb potential. This method is intended to account for the continuum of electron shielding that ranges from the region far from the dust grain where $n_e \approx n_i$ to the region of electron depletion near the dust grains and has been found to reasonably reproduce ion-dust interparticle forces calculated from PIC simulations \cite{box_note_1958}. Further details about the DRIAD simulation and capabilities can be found in \cite{matthews_dust_2020}.
	
	The motion of a ``superion'' with the same charge-to-mass ratio (and consequently, dynamics) as an individual ion is determined by the Yukawa force due to other ions $\vec{F}_{ij}$, the Coulomb force due to charged dust grains $\vec{F}_{iD}$, the alternating axial electric field $\vec{F}_E(x)$, ion-neutral collisions $\vec{F}_{in}$, and boundary conditions imposed by the presumed homogeneous distribution of ions outside the simulation region $\vec{F}_{bound}$. These boundary conditions evolve as the plasma conditions change in the presence of ionization waves, as shown in Fig. \ref{fig:ionwaveparam}. Accordingly, the equation of motion for a superion with mass $m_i$ and charge $q_i$ is given by
	\begin{equation} 
		m_i \ddot{\vec{r}} = \vec{F}_{ij} + \vec{F}_{iD} + \vec{F}_E(x) + \vec{F}_{bound}(r,x,t) + \vec{F}_{in} \label{eq:ionEqMotion}
	\end{equation}
	One superion typically represents approximately 200 individual ions, with the exact number of ions per superion set by the user.
	
	Dust grains with mass $m_d$ and charge $Q_D$ move in accordance with the equation of motion
	\begin{equation}
		m_d \ddot{\vec{r}} = \vec{F}_{di} + \vec{F}_{dD} + \vec{F}_E(x) + \vec{F}_{drag} + \vec{F}_{therm} + \vec{F}_{C}
        \label{eq:dusteqmotion}\end{equation}
	Here, $\vec{F}_{di}$ is the Yukawa force due to the ions, $\vec{F}_{dD}$ is the Coulomb force between dust grains, $\vec{F}_{E}(x)$ is the force of the axial electric field, $\vec{F}_{drag}$ is the neutral drag force (which depends on the temperature and pressure of the neutral gas), and $\vec{F}_{therm} = \zeta R(t)$ is the effect of the dust immersed in a thermal bath. $\vec{F}_C$ is a force that accounts for the dust confinement due to the particular experimental conditions being simulated, including the radial confinement imposed by neighboring dust particles outside the simulation region. The radial component of this confinement is given by 
	\begin{equation} 
		\vec{F} _{C,r} = \omega_1 Q_d \vec{r}_d
		\label{eq:radialConf}
	\end{equation}
	
	where $\omega_1$ is the frequency of the harmonic potential well used to confine the particles in the radial direction and $\vec{r}_d$ is the radial position of the dust. The axial component of the confinement force is given by 
    
	\begin{equation} 
		\vec{F} _{C,x} = \omega_2 Q_d (|x_d| - x_o) \hat{x},
		\label{eq:axialConf1}
	\end{equation}

where $\omega_2$ is the frequency of the confining harmonic potential well in the axial direction. This force acts on particles whose axial position $x_d$ is greater than a given distance $x_o$ from the center of the cylindrical simulation region ($|x_d|>x_o$).  This confines the dust to a region far from the simulation boundaries so that ions re-injected at the boundaries have reached their equilibrium flow speeds in the region of the dust cloud.  
	
\begin{figure}[hbt]
		\centering
		\includegraphics[width=0.42\textwidth]{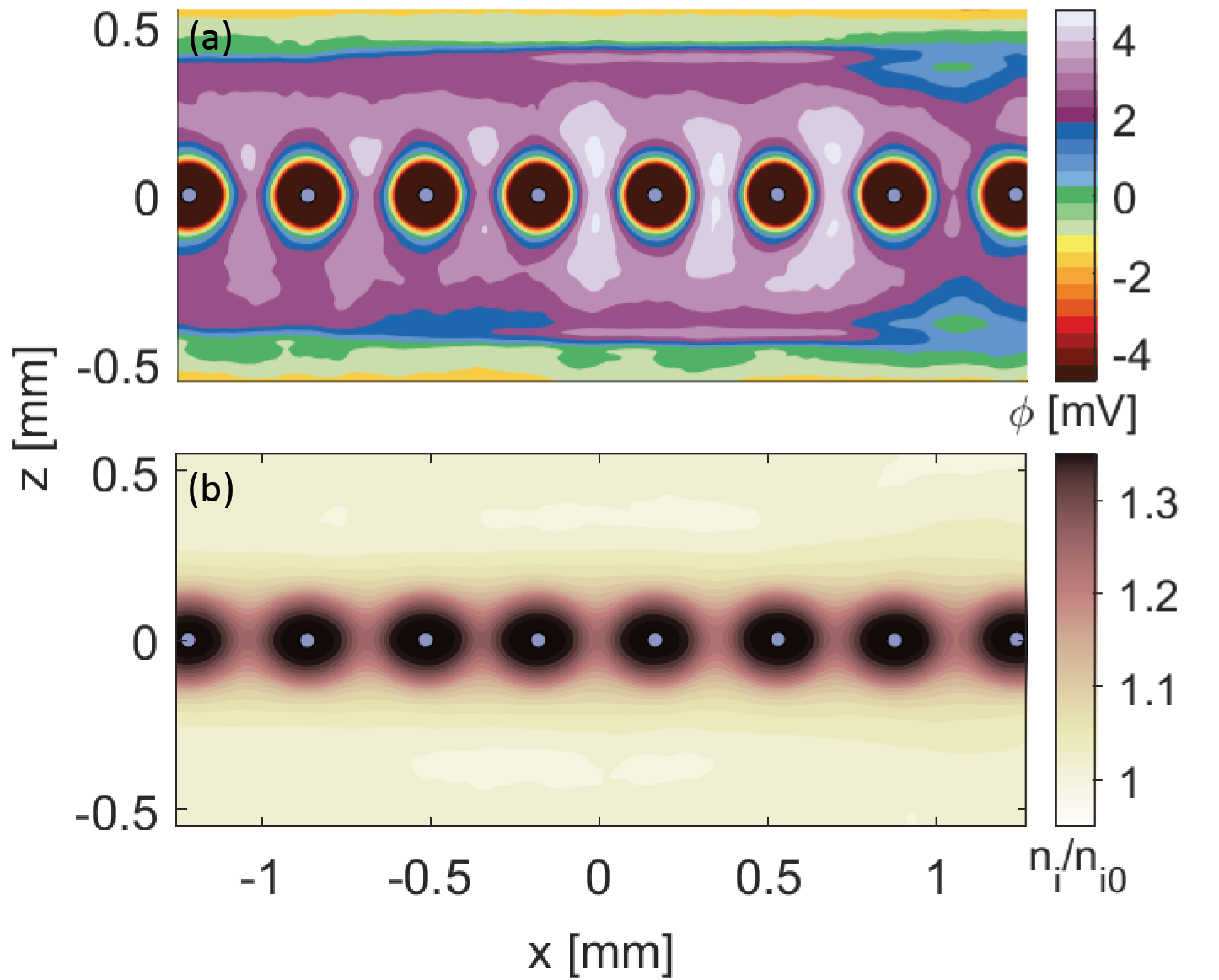}
		\caption{Contour plots of the total electric potential (top) and normalized ion density (bottom) obtained from a DRIAD simulation of 20 dust particles averaged over 0.1 s. Dust positions are indicated by gray circles.  Input parameters for the simulation were  neon gas at 60 Pa with 2.0 mA discharge current, based on the plasma conditions present during the peak axial electric field in ionization waves.}
		\label{fig:densev}
	\end{figure}

	As an example, a simulation of a single filament consisting of 20 dust particles is shown in Fig.~\ref{fig:densev}, where gray circles indicate dust grains.  The total electric potential, calculated as the sum of the ion and dust potentials, is shown in the top panel. The ion density (normalized by the background ion density $4.43 \times 10^{14} ~\mathrm{m}^{-3} $) is shown in the bottom panel. The plasma variations occurring on the microsecond timescale, such as ionization waves, have a significant influence on the ordering of dust structures. As the ionization waves pass, the finite charging and discharging time produces a lower average dust charge than the one calculated from time-averaged plasma conditions. This leads to a weaker repulsive force among dust grains and smaller interparticle spacing \cite{vermillion_influence_2022}. 
	
	\section{Numerical Results}
	
\begin{figure}[h!]
		\centering
		\includegraphics[width=0.45\textwidth]{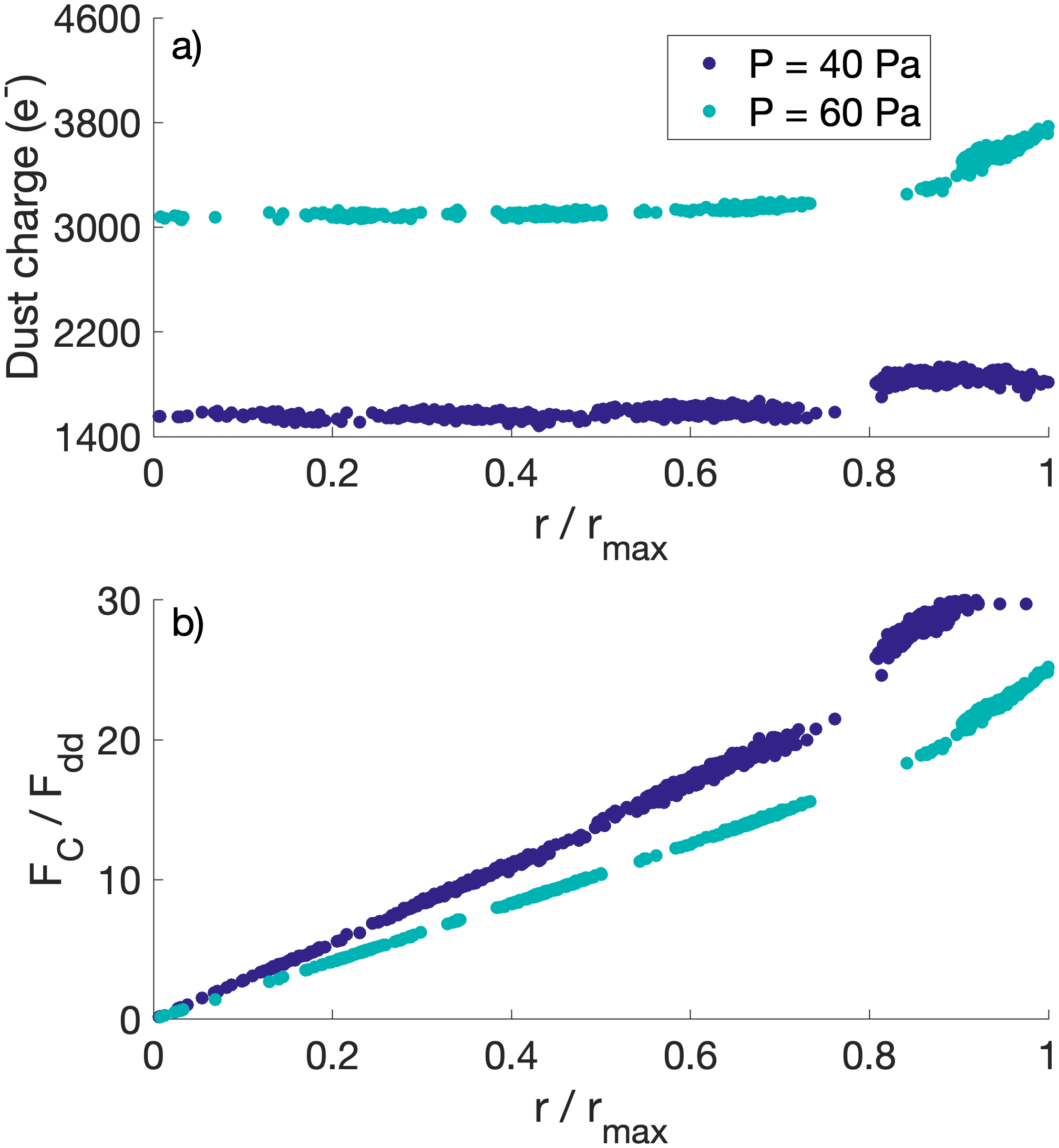}
		\caption{a) Equilibrium charge and b) radial confinement force for the dust grains in the dust clouds for both 40 Pa (purple) and 60 Pa (teal) as a function of the radial dust position within the cylindrical simulation region. The confinement force in b) is normalized by the dust-dust force between two grains with the average charge of grains in inner rings separated by the corresponding average interparticle distance. The distance on the horizontal axis is normalized by the radial position of the outermost dust grain on each case (552  $\mu m$ for 40 Pa and 775 $\mu m$ for 60 Pa).}
		\label{fig:radial_confinement_and_charge}
	\end{figure}

\begin{figure}[h!]
		\centering
		\includegraphics[width=0.45\textwidth]{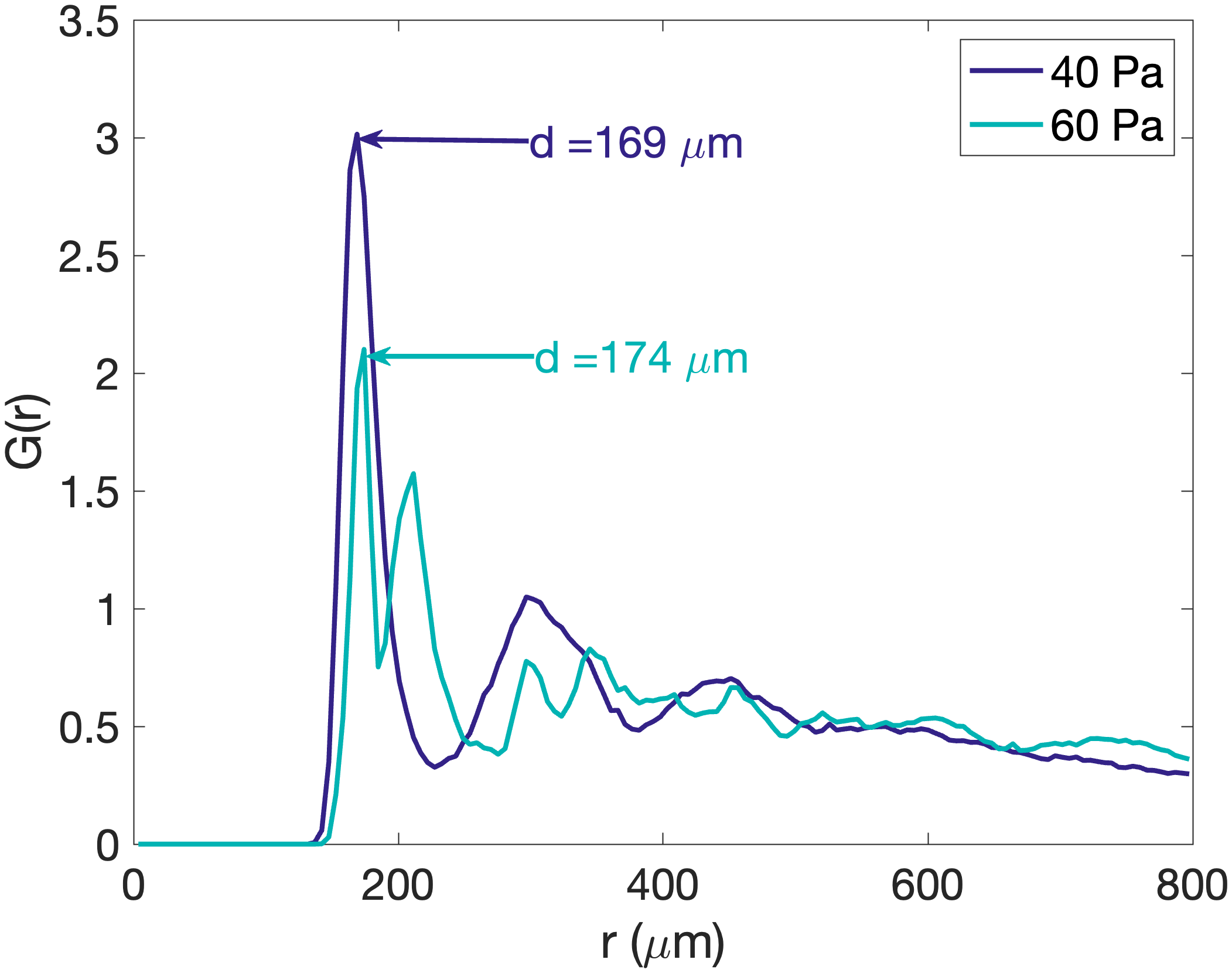}
		\caption{Plots of the pair correlation function $G(r)$ calculated for the simulation results at 40 Pa (purple) and 60 Pa (teal), where both cases involve repeated ionization waves.}
		\label{fig:numerical_gr}
	\end{figure}

	The DRIAD simulation was used to model the dynamics and charging of extended dust clouds under evolving plasma conditions caused by ionization waves for two different conditions: 1) gas pressure 40~Pa and discharge current $I$ = 0.8 mA and 2) gas pressure 60~Pa and discharge current $I$ = 2.0 mA, with an example of the varying conditions shown in Fig.~\ref{fig:ionwaveparam} \cite{hartmann_ionization_2020}. These conditions were selected for two reasons. First, the ionization waves in PK-4 neon plasma have been numerically predicted and experimentally observed at 60~Pa. Therefore, using these conditions allows us to investigate the hypothesis that the observed transition to a liquid crystalline state at higher pressures results from the spiked electric field and plasma response due to the ionization waves. Second, the onset of self-excited dust density waves in PK-4 neon discharge has been observed at 40~Pa when the dust cloud is moved using the electric field \cite{jaiswal2018dust}. The effect is believed to be caused by an ion streaming instability associated with a strong electric field. Therefore, the 40~Pa case allows us to investigate a distinct regime of stability for these dusty plasmas.
	Each simulation modeled 1700 dust grains in a cylindrical region with boundary conditions imposed to confine dust grains at least 2$\lambda_D$ from the simulation boundary in the axial direction and 1$\lambda_D$ from the boundary in the radial direction. We initially adjusted the radial confinement $\omega_1$ to match the interparticle spacing seen in the experiments, and then increased $\omega_2$ (confinement in the axial direction) until crystallization was achieved. The dust grain positions and charges were allowed to evolve to equilibrium. The dust charge at equilibrium and the normalized radial confinement force are shown in Fig. \ref{fig:radial_confinement_and_charge}. As shown in Fig.~\ref{fig:radial_confinement_and_charge}a, the dust charge is relatively uniform in the interior of the dust cloud, but the outermost dust layer has a greater charge as the ion wake is less focused at the exterior of the cloud.
    The radial confinement force is normalized by $F_{dd} = (1/4\pi\epsilon_0)(Q_{eq}^2/d^2)$, where $Q_{eq}$ is the average equilibrium charge for the interior of the dust cloud ($1593e^-$ for 40 Pa and $3123e^-$ for 60 Pa) and $d$ is the average interparticle separation distance, found from the first peak in the plot of g(r) (Fig. \ref{fig:numerical_gr}).
    
    The lowest dust charge was obtained for the lowest pressure (40 Pa) as shown in Fig. \ref{fig:radial_confinement_and_charge}a). 

        \begin{table}[ht!]
		\begin{center}
		\caption{Parameters used in the simulation for both pressures.} 		
		\label{simulation_params}
		  \begin{tabular}{|c|c|c|}
		  \hline
		                                     &  40 Pa  & 60 Pa   \\	
            \hline
            $\omega_1 (\times10^5 N/C m)$	 &  40  & 40  \\	 
            \hline
		    $\omega_2 (\times10^5 N/C m)$	 &  60	 & 60	   \\	 
            \hline
            $<T_i> (\times 10^2 K)$	         &  3.76 & 5.04	   \\	 
            \hline
            $<T_e> (\times 10^2 K)$	         &  739	 & 500	   \\ 
            \hline
            $<n_i> (\times 10^{14} m^{-3})$	 &  2.61 & 12.6	   \\ 
            \hline
            $<n_e> (\times 10^{14} m^{-3})$	 &  2.39 & 12.5	   \\ 
            \hline
		\end{tabular}
		\end{center}
	   \end{table}  
        
\begin{figure}[h!]
		\centering
		\includegraphics[width=0.48\textwidth]{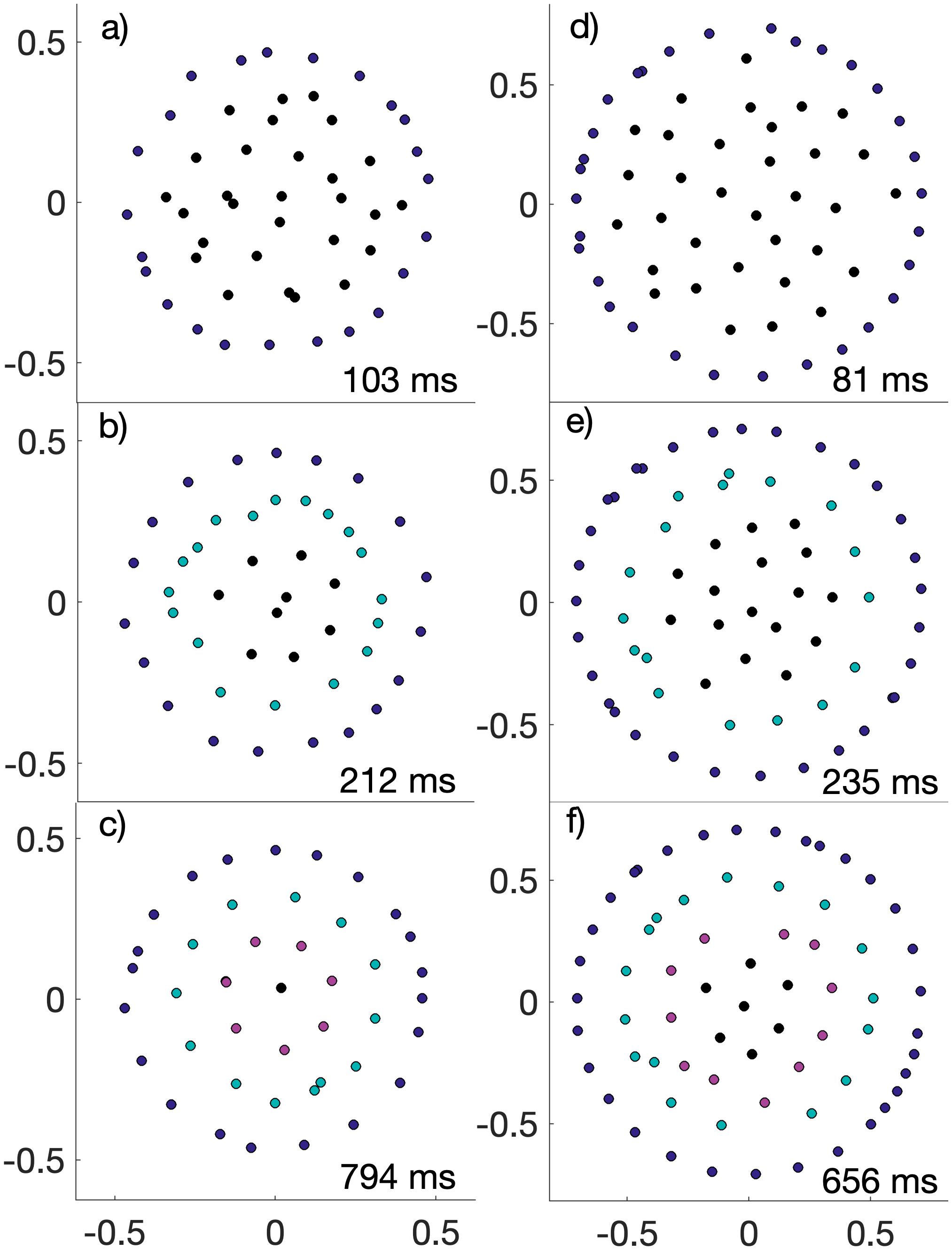}
		\caption{Illustration of the cylindrical layers formation process of a $200 \mu m$ slice of the cloud, with the external layer forming first, followed progressively by inner layers. a-c) 40 Pa, d-f) 60 Pa. The panels show snapshots of different simulation times increasing from top to bottom.}
		\label{fig:layer_formation}
	\end{figure}

 The simulation revealed that the formation of nested cylindrical layers and subsequent crystallization proceeds from the outside in, as shown in the sequence of images in Fig.~\ref{fig:layer_formation}. In Fig.~\ref{fig:dustcloud}, we saw similar formation of layers at the exterior of the clouds cross-sections. However, the formation of nested surfaces in the interior of the clouds is not visible in the experimental data, which is likely due to heating effects. As discussed in \cite{mccabe2025experiments}, in these experiments, an extended time (much greater than the Epstein drag decay) is required to dissipate energy gained at the onset of polarity switching, which means that the clouds probably have not settled into an equilibrium state. The non-equilibrium dynamics of the PK-4 clouds in these experiments was further confirmed by the statistical analysis in \cite{andrew2024anisotropic}.

\begin{figure*}[t]
		\centering
		\includegraphics[width=0.9\textwidth]{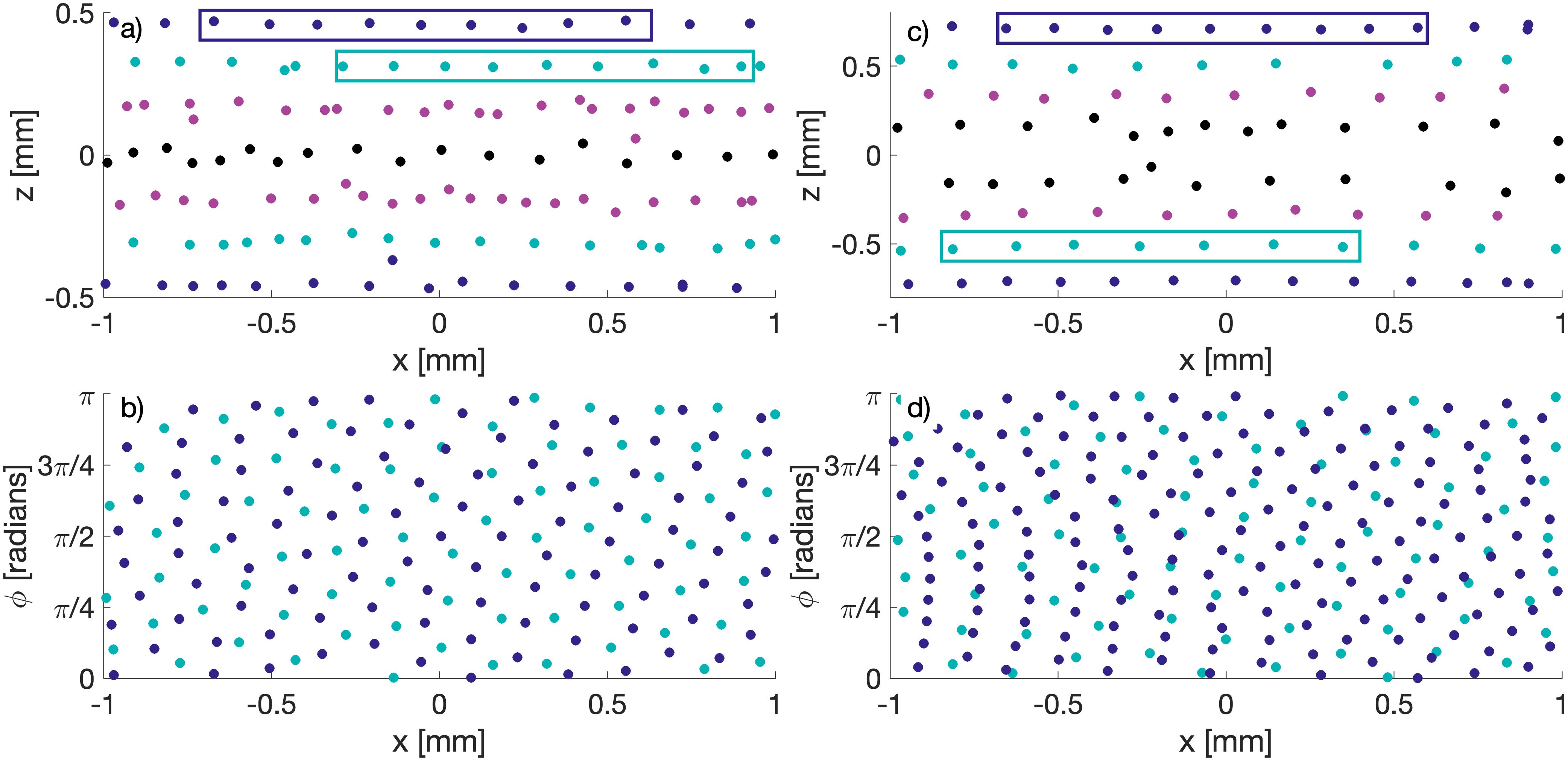}
		\caption{Results from a simulation of a cloud of 1700 dust grains in the PK-4 at: a, b) 40 Pa and c, d) 60 Pa. a, c) 200-$\mu m$ thick slice through the mid-plane showing the arrangement of grains in filamentary structures; c, d) dust positions when the outer two cylinders are 'unwrapped'. Note that only half of the cylindrical shells ($0<\phi<\pi$) are shown. Different colored dots on this figure represent dust particles that belong to the different nested surfaces shown in Fig. 17.}
		\label{fig:dust_cloud_plots}
	\end{figure*}

Similar to experiment, the dust clouds obtained with the DRIAD simulation show the formation of 
dust particle filaments along the axial direction for both pressure cases (Fig.~\ref{fig:dust_cloud_plots}a, c).
We further examine the correlation between the positions of dust particles in adjacent cylindrical layers, which can be seen when the cylinders are ``unwrapped'' by plotting the angular position in cylindrical coordinates ($\phi$) of the dust grains in the nested cylinder vs. the axial position ($x$), as shown in Fig.~\ref{fig:dust_cloud_plots}b,d. 

Comparison of the locations of the particle positions in two adjacent cylindrical shells shows that in the 40~Pa case the particles in one layer tend to be located between the particles in the adjacent layer, which suggests a tendency toward an ordered three-dimensional crystalline arrangement. However, because the number of particles in the two adjacent layers is not the same, the structure is frustrated. This correlated ordering between layers is not seen in the 60~Pa case, which indicates the layers have a higher degree of freedom to move with respect to each other.  This is an illustration of enhanced asymmetric coupling in the system at higher pressures, which is in agreement with the observations from experiment (Fig.~\ref{fig:gphi}).

\begin{figure}[ht!]
		\centering
		\includegraphics[width=0.4\textwidth]{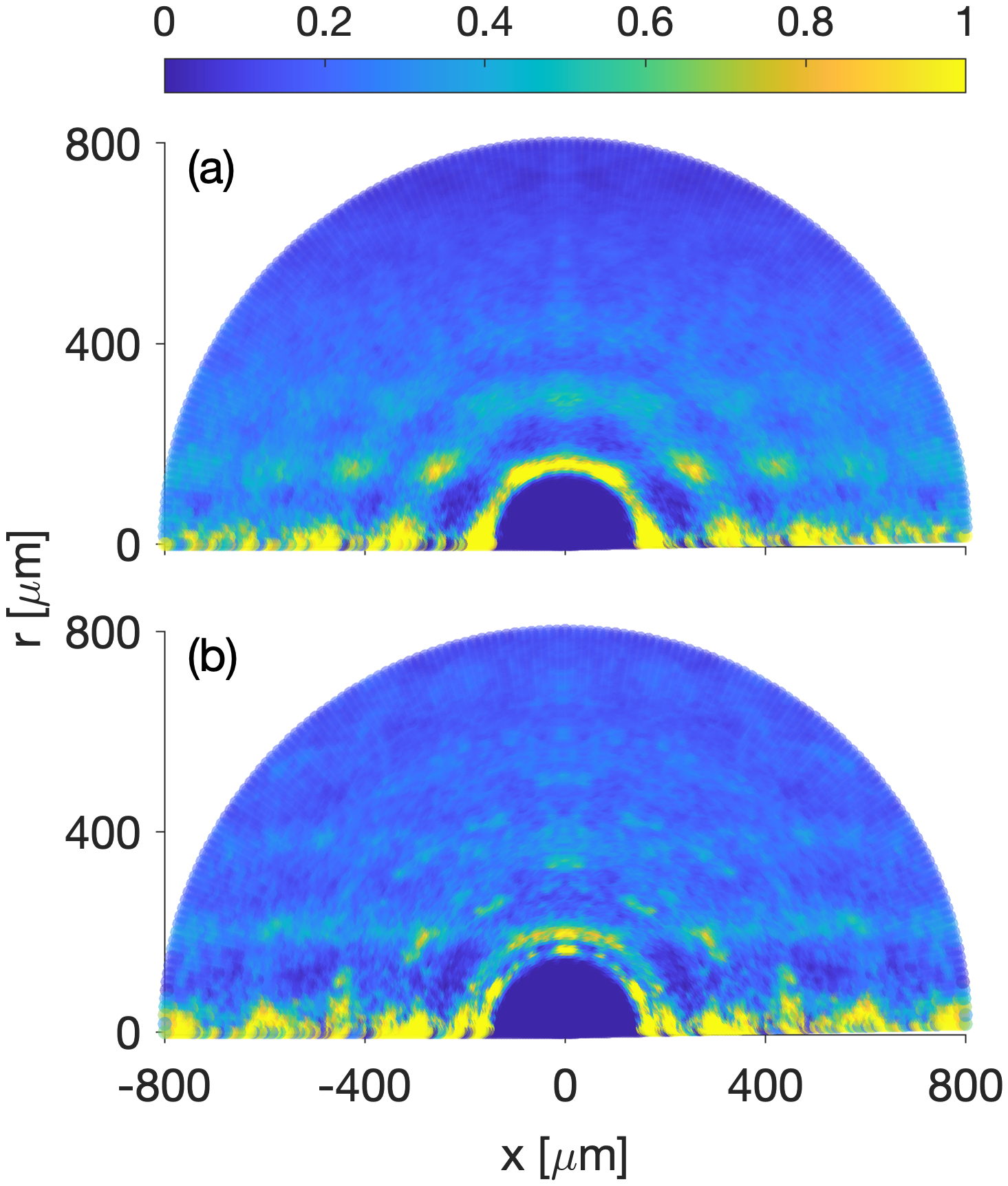}
		\caption{Plots of $G_\phi (r,\theta)$ for the two numerical simulations at (a) 40 Pa and (b) 60 Pa, both conducted with repeated ionization waves.}
		\label{fig:numerical_gphi}
	\end{figure}

\begin{figure}[ht!]
		\centering
		\includegraphics[width=0.35\textwidth]{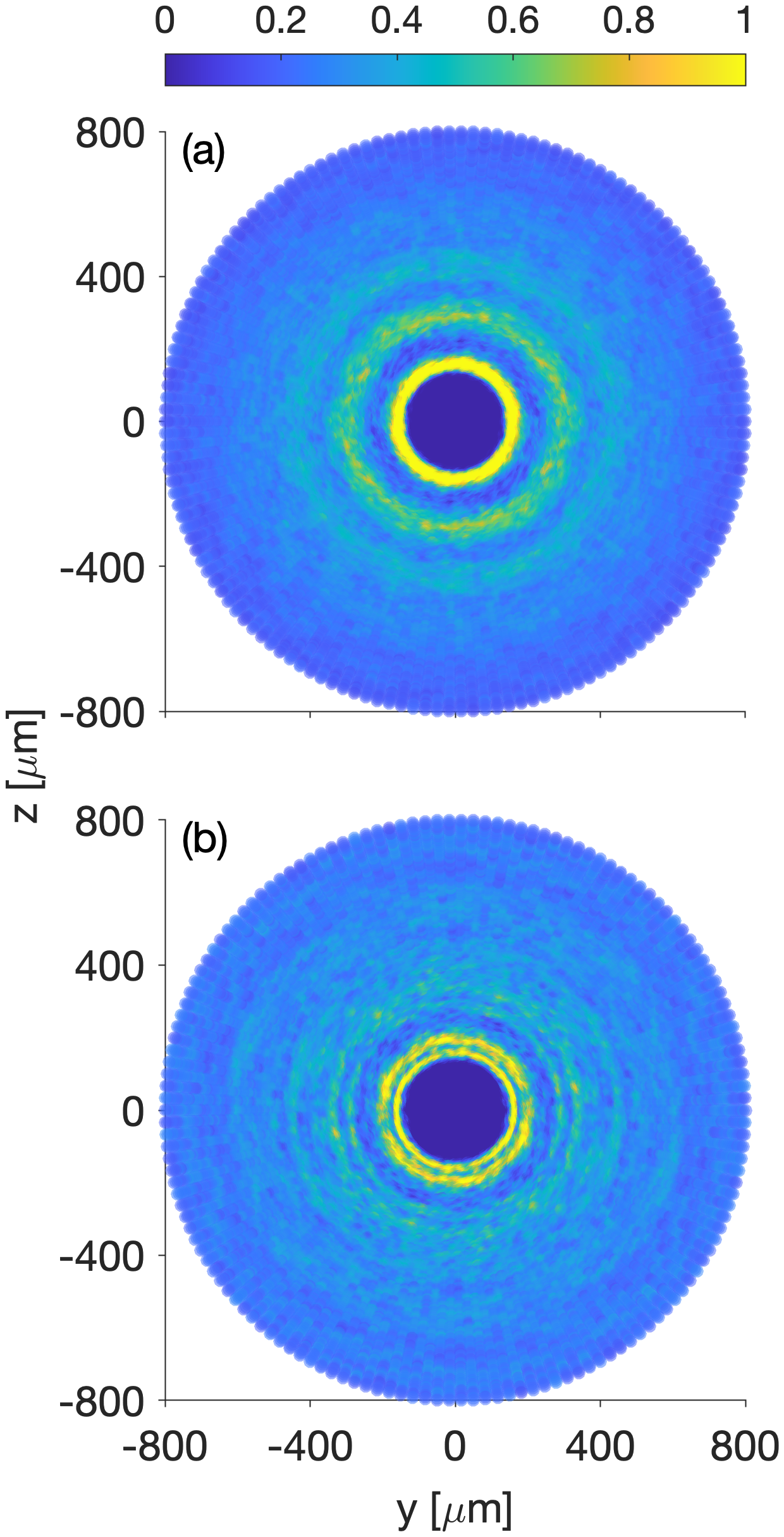}
		\caption{Plots of $G_\theta (r,\phi)$ for the two numerical simulations at (a) 40 Pa and (b) 60 Pa, both conducted with repeated ionization waves.}
		\label{fig:numerical_gtheta}
	\end{figure}

	The 3D structure of the dust clouds was analyzed using the pair correlation functions $G_\phi$ and $G_\theta$, as shown in Figs.~\ref{fig:numerical_gphi} and \ref{fig:numerical_gtheta}, respectively. The string peaks (bright yellow regions along the x-axis) seen in  Fig.~\ref{fig:numerical_gphi}  indicate regular interparticle spacing within filaments. The string peaks are less localized in the 60 Pa case (Fig.~\ref{fig:numerical_gphi}b) than for the lower-pressure 40 Pa case (Fig.~\ref{fig:numerical_gphi}a). The highest crystallinity and degree of order within filaments are observed in the 40 Pa case  (Fig.~\ref{fig:numerical_gphi}a) where the hexagonal symmetry is clearly visible and the string peaks are reasonably well-defined in both sides. The distinct bright spots in the second and third layers reflect the ordering of the particles in adjacent cylindrical shells, as shown in Fig.~\ref{fig:dust_cloud_plots}. The second layer peaks are more diffuse in the 60 Pa case (Fig.~\ref{fig:numerical_gphi}b). This is in agreement with the experimental results where the order in the layered structure formation decreases as the pressure increases (Fig. \ref{fig:gphi}). 
	
The $G_\theta$ results indicate that the 40~Pa case (Fig.~\ref{fig:numerical_gtheta}a) possesses the largest degree of order in the layered structure, indicated by the bright peaks in the second and third concentric rings. In the 60~Pa case (Fig.~\ref{fig:numerical_gtheta}b) $G_\theta$ shows mostly diffuse rings due to filaments along the axial direction not having a preferred location with respect to each other. 
	
	The pair correlation function $G(r)$ calculated for the full cloud in each of the two simulations is shown in Fig.~\ref{fig:numerical_gr}.  The 40~Pa case shows clearer peaks than the 60~Pa case, indicating a higher degree of order within a filament at this pressure. The average interparticle separation (interpreted as the location of the first peak in the $G(r)$ plots) is very similar for both pressures, which is in good agreement with the experimental results.

\section{Conclusion}

Here we presented a structural analysis of dusty plasma experiments from the PK-4 facility on board the International Space Station. Due to the polarity-switched electric field in the PK-4 DC discharge, the dusty plasmas are observed to organize into field-aligned filamentary structures. We observe that, in addition to common orientation along the direction of the external electric field (typical of the electrorheological state), the filaments in PK-4 can also exhibit pressure-dependent pattern formation and layering, which are characteristic of the nematic and smectic states of liquid crystals. To quantify the dust structure for different pressure-current conditions, we calculated various pair correlation functions for the dust particles and compared the results against N-body simulations using the DRIAD code.

We found that, at the lowest experimental pressure, $28.5~\mathrm{Pa}$, the bulk cloud exhibits crystalline properties with similar short-range coupling of particles within filaments and among neighboring filaments. In contrast, at higher pressure ($70.5~\mathrm{Pa}$), the coupling of particles within filaments becomes both stronger and longer-range (enhanced crystalline behavior), while the coupling of particles across filaments decreases. The latter effect indicates an increased degree of freedom for filaments to move with respect to each other (liquid behavior). Since the dust particles in these experiments are suspended in low-temperature plasma, increasing the neutral gas pressure results in ``cooling'' of the dust particles through increased frequency of dust-neutral collisions. The structural changes observed with increasing pressure can then be considered analogous to a transition to a liquid crystal (LC) state as the temperature is decreased. We further observed that the filamentary structures in the experiment exhibit alignment in nested surfaces for several pressure-current conditions, suggesting the possibility of a smectic LC state.

To provide insight into the mechanism causing the observed structural transitions, we conducted numerical simulations of the dust and ion dynamics using the DRIAD code with dynamically evolving plasma conditions obtained from a PIC simulation of the PK-4 discharge. Those conditions model the effects of ionization waves on the equilibrium dust structure. In agreement with experiment, the simulations confirmed the formation of dust filaments at both higher and lower pressure cases. The simulations also reveal that the dust particles tend to organize into nested cylinders aligned with the axial electric field direction, with the ordering proceeding from the outermost to the innermost cylinder. Mapping the dust positions within neighboring nested surfaces showed a tendency for crystalline alignment at lower pressures. At higher pressures, the particles within neighboring surfaces exhibit enhanced degree of freedom to move, suggesting decreased cross-surface coupling. This confirms the experimental observation of non-isotropic coupling at higher pressures. We further confirm that at higher pressure, both the strength and range of coupling within filaments is enhanced while the cross-filament coupling is decreased. 



These results provide evidence that microgravity dusty plasmas exhibit a rich variety of LC-like structure, including field alignment, layering, and pattern formation. Based on these results, we propose that microgravity dusty plasma experiments can be used as a macroscopic analogue system for the study of fundamental questions in the theory of liquid crystals, such as the universality of LC phase transitions, the origins of pattern formation, and the propagation of defects at critical conditions. 

\section{Acknowledgments}
This material is based on work supported by NSF grant numbers 2308742, 2308743, 1903450 and 1740203, NSF EPSCoR FTPP OIA2148653, NASA JPL 1571701, and US Department of Energy, Office of Science, Office of Fusion Energy Sciences under award number DE-SC0021334. All authors gratefully acknowledge the joint ESA - Roscosmos ``Experiment Plasmakristall-4'' onboard the International Space Station. The microgravity research is funded by the space administration of the Deutsches Zentrum für Luft- und Raumfahrt eV with funds from the federal ministry for economy and technology according to a resolution of the Deutscher Bundestag under Grants No. 50WM1441 and No. 50WM2044. P. Hartmann acknowledges the support by the National Office for Research, Development and Innovation (NKFIH) via Grant K134462.

\bibliography{new_bib}

\end{document}